\documentclass[journal]{IEEEtran}
\usepackage{setspace}
\usepackage{multicol}
\usepackage{stfloats}
\usepackage{amsmath}
\usepackage{titlesec}
\usepackage{mdframed}
\usepackage{amsmath}
\usepackage{subcaption}
\usepackage{float}
\usepackage{enumitem}

\usepackage{titlesec}

\titleformat{\section}{\bfseries\centering}{\thesection}{1em}{}
\titleformat{\subsection}{\bfseries\centering}{\thesubsection}{1em}{}

\makeatletter
\renewcommand{\@seccntformat}[1]{%
    \ifcsname the#1\endcsname
        \csname the#1\endcsname\ 
    \fi
}

\renewcommand{\thesection}{\arabic{section}}
\renewcommand{\thesubsection}{\thesection.\arabic{subsection}}

\makeatother
\usepackage{graphicx}
\usepackage{caption} 
\graphicspath{ {./images/} }
\usepackage[a4paper, margin=1in]{geometry}
\usepackage{xcolor} 
\usepackage{listings} 
\usepackage{times}
\fontsize{10}{12}\selectfont 
\usepackage{amsmath}
\usepackage{cuted}

\ifCLASSINFOpdf

\else

\fi

\colorlet{punct}{red!60!black}
\definecolor{background}{HTML}{EEEEEE}
\definecolor{delim}{RGB}{20,105,176}
\colorlet{numb}{magenta!60!black}

\lstdefinelanguage{json}{
    basicstyle=\scriptsize\ttfamily,
    numbers=left,
    numberstyle=\tiny,
    stepnumber=1,
    numbersep=8pt,
    showstringspaces=false,
    breaklines=true,
    frame=lines,
    backgroundcolor=\color{background},
    literate=
     *{0}{{{\color{numb}0}}}{1}
      {1}{{{\color{numb}1}}}{1}
      {2}{{{\color{numb}2}}}{1}
      {3}{{{\color{numb}3}}}{1}
      {4}{{{\color{numb}4}}}{1}
      {5}{{{\color{numb}5}}}{1}
      {6}{{{\color{numb}6}}}{1}
      {7}{{{\color{numb}7}}}{1}
      {8}{{{\color{numb}8}}}{1}
      {9}{{{\color{numb}9}}}{1}
      {:}{{{\color{punct}{:}}}}{1}
      {,}{{{\color{punct}{,}}}}{1}
      {\{}{{{\color{delim}{\{}}}}{1}
      {\}}{{{\color{delim}{\}}}}}{1}
      {[}{{{\color{delim}{[}}}}{1}
      {]}{{{\color{delim}{]}}}}{1},
}
\lstdefinelanguage{JavaScript}{
    keywords={break, case, catch, continue, debugger, default, delete, do, else, finally, for, function, if, in, instanceof, new, return, switch, this, throw, try, typeof, var, void, while, with, let, const, async, await, yield, class, extends, constructor, static, super},
    ndkeywords={true, false, null, undefined, NaN, Infinity, console, window, document, Math, Array, Object, String, Number, Boolean},
    sensitive=true,
    comment=[l]{//},
    morecomment=[s]{/*}{*/},
    morestring=[b]",
    morestring=[b]',
    morestring=[b]`,
    basicstyle=\ttfamily\footnotesize,
    keywordstyle=\color{blue}\bfseries,
    ndkeywordstyle=\color{teal},
    stringstyle=\color{red},
    commentstyle=\color{green!50!black},
    numbers=left,
    numberstyle=\tiny\color{gray},
    stepnumber=1,
    backgroundcolor=\color{lightgray},
    frame=lines,
    breaklines=true,
    literate=
     *{:}{{\color{black}:}}{1}
      {,}{{\color{black},}}{1}
}

\definecolor{lightgray}{rgb}{0.95,0.95,0.95}

\usepackage{setspace}
\usepackage[hidelinks]{hyperref}
\begin{document}
\onehalfspacing

\title{AI-Driven Electronic Health Records System for Enhancing Patient Data Management and Diagnostic Support in Egypt}

\author{
    \IEEEauthorblockN{
        Arwa Alorbany\textsuperscript{1}, 
        Mariam Sheta\textsuperscript{1}, 
        Ahmed Hagag\textsuperscript{1}, 
        Mohamed Elshaarawy\textsuperscript{1}, 
        Youssef Elharty\textsuperscript{1}, 
        and Ahmed Fares\textsuperscript{1}\\
    }
    \IEEEauthorblockA{
        \textsuperscript{1}Department of Computer Science and Engineering, \\
        Egypt-Japan University of Science and Technology\ \\
        Email: \{arwa.zakaria, maryem.abousaad, ahmed.hagag, mohamed.elshaarawy, youssef.elharty, ahmed.fares\}@ejust.edu.eg
    }
}

\maketitle

\begin{abstract}
The digitalization of healthcare infrastructure is essential for medical service delivery and improving people lives globally. However, Egypt faces significant barriers to adopting Electronic Health Records (EHR), with only 314 hospitals implementing such systems as of October 2024. This slow adoption limits efficient data management and impedes healthcare providers from making timely and accurate decisions. To address these challenges, this project introduces an EHR system specifically designed for Egypt’s Universal Health Insurance and the healthcare ecosystem overall. The proposed system aims to simplify patient data management by centralizing medical histories with scalable micro-services software architecture and a polyglot persistence strategy to facilitate real-time access to patient information, and improving communication between healthcare providers. The system also enhances clinical workflows by integrating patient examinations and medical history tracking. At its core, the system features the Llama3-OpenBioLLM-70B model, a state-of-the-art biomedical language model, to generate actionable summaries of patients' medical histories, provide chatbot feature, and automatically generate a medical report with AI recommendations. This enables healthcare providers and doctors to perform efficient searches and prompts for specific information during consultations. Additionally, Vision Transformer (ViT) model is used for pneumonia disease classification to assist radiologists in reporting. Quantitative evaluations using ROUGE and BERTScore demonstrate that while the AI model excels in capturing essential medical details (high recall), improvements are needed in generating concise, coherent narratives.  The study’s findings imply that with further optimization—such as incorporating retrieval-augmented generation, fine-tuning on local data, and integrating standardized interoperability protocols—the proposed AI-driven EHR system could significantly enhance diagnostic support, clinical decision-making, and overall healthcare delivery in Egypt.

\end{abstract}


\IEEEpeerreviewmaketitle

\section{Introduction}

\IEEEPARstart
Healthcare is a foundational need for humankind. It affects our opportunity to pursue life goals, reduce our pain and suffering, helps prevent premature loss of life, and provides information needed to plan for our lives. The community has an obligation to provide medical care with high quality for all its members. The need to deliver efficient and high-quality services has put the healthcare providers on the spotlight. In Egypt, these pressures are compounded systemic inefficiencies in patient data management and healthcare delivery. This retards the ability of providers to offer timely and effective care for patients. The following obstacles highlight the gaps in Egypt’s current healthcare infrastructure:
\begin{enumerate}

    \item \textbf{Fragmented Patient Records:} Fragmentation of patient records across multiple medical facilities causes a gap in the consultations. In Egypt, there is no centralized system for tracking or retrieving patient histories, leaving medical data scattered and disorganized. This often results in incomplete or missing information during consultations. As a result, patients are responsible for recalling and providing their own medical history. This act increases the risk of misdiagnoses, repeated tests, and delays in treatment. \cite{ehis}
    \item \textbf{Poor Access to Information: }A major challenge is the lack of access to a patient’s medical history. Doctors often struggle to find updated information about their patients. Also, Paper-based and semi-digital systems increase the chances of errors in data entry, storage, and retrieval. These gaps make it harder for doctors to make informed decisions, which can negatively impact treatment and outcomes.
    \item \textbf{Lack of Standardization in Medical Records:} In Egyptian hospitals, medical records are not standardized. Some clinics still use handwritten records, while others rely on digital systems that are not interconnected. This lack of standardization creates serious challenges when patients move between different providers or hospitals. Standardized record-keeping is essential for ensuring continuity of care. It allows patient information to be seamlessly transferred and integrated across the healthcare system.
    \item \textbf{Limited Data-Driven Decision Making:} Data is a cornerstone of modern healthcare. It helps identify trends, predict health risks, and support informed decision-making. However, in systems without proper digital infrastructure, valuable patient information often remains unused. For example, patterns in chronic diseases might go unnoticed, and the effectiveness of treatments may not be properly evaluated. Poor data management also limits the ability to conduct large-scale public health studies or allocate resources efficiently. \cite{ehis2}
\end{enumerate}

\subsection{Problem statement}
\par To serve the approximately 100 million people that reside in the narrow valley area surrounding the Nile River, Egypt recently adopted digital transformation. The country intends to use contemporary medical technologies to provide its citizens with effective and affordable healthcare. \cite{phdthesis} \\
The amount of data produced by millions of patients is too much for paper-based systems, which are still in use today. However, there were several limitations to the paper-based records that hampered their quality, such as unclear doctors’ notes, which were often misinterpreted and may lead to significant clinical mistakes by pharmacists \cite{hatton}. These traditional record-keeping methods also suffer from limited accessibility and security concerns, as paper records are exposed to loss, damage, and unauthorized access, further complicating healthcare delivery.\\
Based on embracing a vision of digital transformation and using information technology to enhance decision-making as part of Egypt's Vision 2030, the Egypt Healthcare Authority unveiled the main facets of its telemedicine and digital transformation projects. \cite{eha-highlights-transformation} The use of electronic health records (EHRs) in public hospitals and clinics is one of the initiatives the government has started to update the healthcare system. The adoption rate is still uneven, nevertheless, with private healthcare facilities and urban areas spearheading the shift while public and rural healthcare facilities lag behind because of infrastructural issues and resource limitations.\cite{BADRAN2019576} This lack of widespread adoption can be attributed to several interconnected challenges. \cite{article}\\
Additionally, the lack of AI-driven insights for doctors in existing systems limits their ability to make data-informed decisions, which could otherwise improve diagnostic accuracy, treatment plans, and patient outcomes.
\par Existing EHR systems in Egypt are often not designed to handle the scale of data processing required for a population of this size. This results in system slowdowns, crashes, and data bottlenecks. The lack of advanced data analytics capabilities further limits the ability of healthcare providers to derive meaningful insights from patient data, hindering efforts to improve care delivery and public health outcomes.
Moreover, patient engagement in their medical history is minimal, as current systems do not provide accessible platforms for patients to view or manage their health information, reducing their ability to participate actively in their care.
The major outbreak of Covid-19 has highlighted the need for healthcare organizations to communicate in a common language to address the challenges of this global crisis. The rapid rise in cases over such a brief time has increased the pressure on health systems to take urgent actions.\cite{article5}
\par In this context, an integrated AI EHR can overcome some of the challenges related to data management, predictive analytics, and real-time decision-making. It can be made more accessible on cloud platforms, secure with advanced encryption of data, and more efficient by automating routine tasks, reducing manual errors, and enhancing productivity. Furthermore, with AI-driven insights, doctors will be empowered to spot trends and foresee outbreaks while making treatments more personalized. The patient portals also increase engagement with easy access to full medical history, progress tracking, and the ability to communicate with healthcare providers.
Security, accessibility, and efficiency in data handling are salient features which cannot be gainsaid. For example, a strong AI-integrated EHR system should ensure data protection from breaches, easy access by authorized users both in rural and urban areas, and easy workflow to handle the large volumes of data resulting from Egypt's huge population. Such a system, if it addresses these critical aspects, would be revolutionary in health care delivery, bridge the urban-rural divides in facilities, and hopefully improve health outcomes for millions of Egyptians.
\subsection{Research objectives}
\begin{itemize}[left=0pt] 
    \item Develop an AI-integrated EHR system tailored for Egypt, considering security concerns, local healthcare needs, regulations and technological barriers.
    \item Conduct a survey to assess Egyptian doctors' and patients' perspectives on EHR, with a focus on usability, trust, and anticipated benefits to find how informative and aware they are and also their opinions on what functionalities would assist them.
    \item Enhance doctor-patient-admin interactions with secure role-based access by implementing secure role-based access control using restricted back-end authorization tools as Auth0 and end-to-end access control to restrict access based on rules ensuring privacy and streamline communication.
    \item Integrate AI functionalities for medical data summarization (e.g., summarizing patient histories) and automated x-ray classification with adding a chatbot in doctor’s views for faster data retrieval, aiming to improve diagnostic accuracy and efficiency.
\end{itemize}
\subsection{Significance of the study}
\par This study would contribute to the Egyptian health care system and its users (health professionals and patients) and would give insights for the future research and serves as a benchmark for developing countries that implement the system. For the healthcare system, it improves efficiency in patient data management, ensures secure and organized record-keeping, and provides one system that links all countries’ health affiliation to the patients’ history. For doctors, it allows easier access to patient history for better and faster diagnosis and provides AI assistance in summarizing records and classifying x-rays. For patients, it offers greater control over their medical records, the ability to request data updates and additions for accuracy, and the ability to request and schedule examinations. For future research, it provides insights into the acceptance of AI-enhanced EHRs in Egypt and can serve as a model for similar implementations in other developing countries, offering lessons learned and best practices.
\subsection{A brief overview of the methodology}
Prior to development, a survey with both healthcare professionals and patients in Egypt took place to understand key requirements and EHR system-related challenges. EHR development follows an agile manner, with six discrete sprints in its structure to achieve the project's milestones. \\
Microservice architecture with key modules such as user management, patient record, and artificial intelligence (AI) integration is utilized in this system. AI capabilities, powered by Llama3 OpenBioLLM-70B and ViT X-ray Pneumonia Classification, contribute towards enhancing medical workflows through medical history summation, reporting, and diagnostics. Security and compliance are prioritized through a multi-dimensional mechanism, including use of Auth0 for authentication, role-based access controls (RBAC), encryption, and logging, in an attempt to comply with Egypt’s Data Protection Law. Deployment of the system involves a containerized model supported through Docker and Kubernetes, supplemented with continuous integration and continuous delivery (CI/CD) pipelines through GitHub Actions, and hosting in Azure Kubernetes Service (AKS) for both scalability and dependability.

\subsection{Paper organization statement}
This thesis is organized as follows, section one introduces the research problem, objectives and the significance of the study. Section two reviews the related work and the previous EHR systems made in Egypt. Section three describes the research methodology from the survey conducting to the system design and architecture. Section four presents the results of the survey, the results of analysis of summarization task by the AI using ROUGE and BERT Score and section five discusses their implications. Section six concludes the thesis and section seven contains the recommendations for future research.
\section{Related Work}
The integration of Electronic Health Record (EHR) systems and Artificial Intelligence (AI) in the healthcare sector transformed the management of patient data. This section is structured around the application of EHR systems in healthcare organizations. In addition, it outlines recent developments and provides a critical analysis of existing systems to form the foundation of the proposed project.
\subsection{Paper-Based Medical Records}
Healthcare systems have long relied on paper-based medical records for the documentation of patient information. Handwritten notations and laboratory tests have been included in such records. However, paper records have significant disadvantages that make healthcare delivery less effective. A study titled "Paper-Based Medical Records"  presents insightful observations regarding such records.\cite{article6}
The study reveals that paper-based medical records have no efficiency in terms of information retrieval. There could be a delay in accessing important information in the case of an emergency as these records have to be stored in physical filing systems.
\subsection{EHR Systems}
Electronic Health Record (EHR) systems have become an integral part of modern-day healthcare organizations, allowing for enhanced management and access to patient information. According to a study in the International Journal of Health Sciences and Research, the use of EHR systems has significantly increased care delivery, particularly in coordination between care providers, reducing medical errors, and enhancing patient outcomes. By transforming patient information into a computerized format, EHR systems have optimized access for medical professionals to critical information, such as medical histories, laboratory tests, imaging tests, and medication lists, all delivered in a timely and efficient manner.\cite{Odekunle2017}\\ 
The use of EHR systems has initiated a significant change in practice in terms of enhancing collaboration and information dissemination in a variety of care settings. EHR system interoperability ensures that critical patient information is accessible to approved care providers whenever and wherever it is needed, minimizing care delivery times and enhancing overall care service quality. According to IJHSR, EHR systems enable better decision-making by providing medical professionals with updated information about a patient's state.
The key EHR system traits include:

\begin{itemize}
    \item \textbf{Centralized Data Management} Storing patient data, such as medical history, laboratory tests, imaging, and medications, in a single digital repository.
    \item \textbf{Interoperability} Allowing for the exchange of information among healthcare providers.
    \item \textbf{Patient-Centered Care} Enabling patients through portals and mobile applications to view their health records and interact with providers.
\end{itemize}
\subsection{AI in EHR Systems}
The integration of AI in clinical workflows is very critical, specifically in Egypt, for effective utilization. Research conducted highlights the role of AI in automating labor-intensive tasks, including analysis and reports generation. It declares the importance of user centric AI tools that adapt to diverse healthcare environments.\cite{article1} 
Public hospitals in Egypt often operate under significant pressure and this integration has the potential to optimize workflow and streamline data access through the organization, especially in emergency cases. Key applications of AI in healthcate include:
\begin{itemize}
    \item \textbf{Clinical Decision Support} Using AI to analyze patient data and provide evidence-based recommendations for diagnosis and treatment.
    \item \textbf{Predictive Analytics} Identifying high-risk patients and predicting disease outbreaks or complications.
\end{itemize}
In the editorial "Artificial Intelligence Algorithms for Healthcare," a thorough analysis is presented concerning the role of AI algorithms in modern-day healthcare. According to the authors, AI-powered approaches are revolutionizing diagnostics, personalized care planning, medical imaging, and administrative efficiency. In addition, they highlight an increased reliance on machine learning (ML), deep learning (DL), and natural language processing (NLP) methodologies, contributing to improvements in several aspects of delivering care, such as the analysis of patient information and predictive model development. \cite{Chumachenko2024}.
Transitioning from a theoretical concept to a critical part of the healthcare industry. According to the authors, AI-powered tools have already proven significant improvements in early disease detection, medical imaging interpretation, and the automation of medical workflows. For instance, medical tools with inbuilt machine learning capabilities can detect cardiovascular disease and cancer with high accuracy. Likewise, algorithms in natural language processing are utilized to process unstructured medical documents, supporting effective information extraction and integration in medical care systems.
Also, AI-powered predictive analysis in electronic medical records enables high-risk patient identification. With such a feature, early interventions and smart distribution of resources can occur. In addition, AI-powered automation in medical billing and claims processing simplifies such processes.
\subsection{ Advances in AI-Based EHR Systems}
Recent advancements in AI have significantly enhanced the capabilities of EHR systems:
\subsubsection{Generative AI for Data Summarization}
Generative artificial intelligence, specifically in terms of data summarization through models based on Retrieval-Augumented Generation (RAG), is increasingly being utilized to generate and interpret complex medical information, and in the process, ease the cognitive burden for clinicians. For instance, in "Enhancing EHR Analysis: Leveraging RAG-Enabled Generative AI for Clinical Data Summarization," such technology is discussed in detail about how it can effectively present important patient information and actionable information.\cite{article3} It highlights how this approach can help: 
\begin{itemize}
    \item \textbf{Summarize patient records} Extracting critical details like diagnosis history, medications, and lab results into concise reports.
    \item \textbf{Reduce physician workload} Automating the summarization of lengthy clinical notes, saving time during consultations.
    \item \textbf{Enhance decision-making} Providing real-time, context-aware insights based on patient data and clinical guidelines.
\end{itemize}
Although RAG maximizes medical information summarization, it can be affected by token restrictions present in the input information. Large language models (LLMs) rely on contextual information derived through documents that have been retrieved; yet, when dimensions in terms of input tokens become restricted, a variety of limitations become apparent:
\begin{itemize}
    \item Loss of Contextual Information
    \item Difficulty in Capturing Long-Term Patient History
    \item Inefficient Multi-Step Processing
\end{itemize}
A study, "Clinical Text Summarization: Adapting Large Language Models Can Outperform Human Experts," considers whether it is feasible for large language models (LLMs) to generate summaries of clinical documents not only at a level equivalent to but even exceeding that of expert clinicians. Inclusion of LLM-enabled summarization tools in EHRs can make it feasible to manage high volumes of clinical information in an efficient manner. By providing concise summaries of patient histories and pertinent information, such AI tools can ease clinicians' cognitive loads and can potentially contribute towards better patient outcomes through optimized decision-making.\cite{Van_Veen2023-cw}
\subsubsection{Priority-Based Medical Segmentation}
Priority-Based Medical Segmentation is a new technique for enhancing both care delivery efficiency and efficacy, particularly in high-pressure environments such as in emergency departments. It involves prioritization and extraction of critical patient information from electronic health records (EHRs), and in doing so, enables quick access to most relevant information for immediate use in clinical decision-making. Research, such as in "Designing an AI Healthcare System: EHR and Priority-Based Medical Segmentation Approach," identifies the importance of such a technique in enhancing clinical workflows and patient care outcomes.\cite{article4}
\subsubsection{Real-Time Analytics in AI-Driven EHR Systems}
The integration of AI in EHRs has significantly enhanced real-time medical evaluation, particularly in radiology. One significant use of such technology is in deploying pretrained AI models for X-ray image analysis, allowing medical professionals to effectively evaluate medical visuals and make data-dependent decisions in an efficient manner.\\
"Leveraging Pretrained Models for Multimodal Medical Image Interpretation: An Exhaustive Experimental Analysis" describes how deep learning models can be utilized to promote multimodal medical image analysis, including X-ray diagnostics, through real-time AI evaluations in a clinical setting.\cite{Fagbola2024.08.09.24311762}\\
Traditional X-ray interpretation is a labor-intensive, time-consuming process for radiologists. However, AI-powered real-time analytics facilitates:
\begin{itemize}
    \item \textbf{Real-time abnormality detection:} AI can signal potential problems (e.g., fractures, pneumonia, or lung nodules) in seconds.
    \item \textbf{Improved workflow productivity:} Real-time X-ray evaluation eliminates reporting delay time, allowing earlier diagnosis and treatment planning.
    \item \textbf{Integration with EHR systems:} AI-produced reports can be easily inserted into an electronic patient record.
\end{itemize}
\subsection{AI Integration in Clinical Workflows}
The integration of artificial intelligence (AI) in clinical workflows has potential for automation of workloads with high labor requirements, improving overall operational efficiency in general. AI technology can generate clinical documentation, freeing medical professionals from clerical workloads. In addition, AI-powered telemedicine platforms enable virtual consultation and continuous follow-up of long-term medical conditions.

\subsection{Electronic Health Record Systems and Scheduling Software in Egypt}

\begin{figure}[h]
    \centering
    \includegraphics[width=1.0\linewidth]{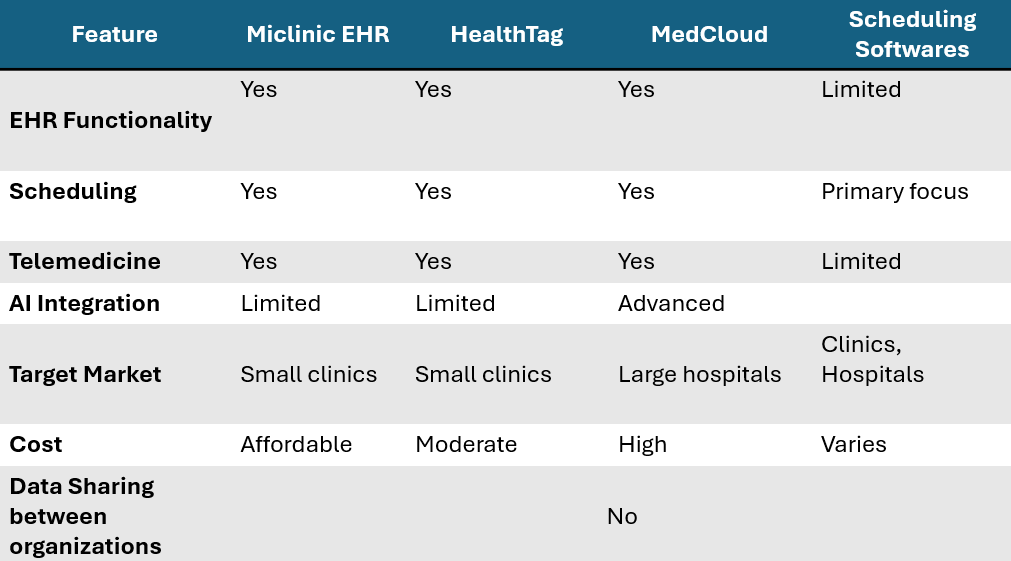}
    \caption{Comparison of EHR applications in Egypt and Scheduling Software Solutions}
    \label{comparison}
\end{figure}

Health care professionals in Egypt increasingly use digital technology to improve operational efficiency and patient satisfaction. Scheduling software is a key tool for clinic workflows and service delivery in terms of service quality improvement. With increased demand for medical care, numerous scheduling programs have been designed specifically for use in countries with specific requirements, including Egypt. Not only can these tools schedule appointments, but they can also integrate with Electronic Health Record (EHR) platforms, providing an integrated platform to provide care to patients. The following table in figure \ref{comparison} shows software platforms in use in Egypt, prevalent in medical practice, and with important roles in the medical field.

The table compares a selection of key Electronic Health Record (EHR) and scheduling software options in the Egyptian marketplace, with a focus on their key features, target segments, and capabilities. Miclinic EHR, HealthTag, and MedCloud stand out as key EHR platforms, each with a specific target in the healthcare sector. 

In particular, Miclinic EHR and HealthTag target small clinics and offer budget options with key capabilities such as scheduling, telemedicine, and basic EHR capabilities. However, artificial intelligence integration is not a key feature in them. 

On the other hand, MedCloud targets big hospitals, offering high-end capabilities such as strong AI integration, high-level telemedicine capabilities, and high scalability. Despite its high price, MedCloud's ability to manage complex networks in healthcare and enable interoperability makes it a preferred tool for larger healthcare providers.

Standalone scheduling software options, in contrast, have a sole focus on scheduling, an integral part of clinic and hospital operations. Scheduling software programs often integrate with EHR platforms to enable a harmonized workflow; however, EHR capabilities in them are limited. The table identifies a lack of interoperability in terms of data sharing between organizations, a critical consideration in modern-day healthcare networks. MedCloud, for instance, promotes high-level data sharing via HL7 and FHIR standards, but alternatives such as Miclinic EHR and HealthTag lack strong capabilities in such a feature.
To address such gaps in Egypt's healthcare technology environment, our project is focused on creating an upgraded EHR system with state-of-the-art artificial intelligence capabilities for enhancing information dissemination, interoperability, and clinical decision-support structures. In particular, the aim is to correct current systems' weaknesses, including the lack of integration of sophisticated AI in budget-friendly alternatives. Our proposed system incorporates AI models for efficient medical information summarization, and in doing so, enables medical professionals to extract actionable and concise information out of lengthy patient files. With such improvements, our project aims to introduce a cost-effective and EHR platform that not only addresses requirements for hospitals but can serve widespread networks, and in the long run, promote patient care and operational efficiency in Egypt's medical sector.

\section{Methodology}
This section, informed by a survey conducted among healthcare professionals and end-users, details the design, architecture, and implementation of the Electronic Health Record (EHR) system, along with its core functionalities.It outlines the software development lifecycle (SDLC), emphasizing the Agile methodology as the framework for delivering a scalable and secure system, the specific components of the system and explains why each technology stack within the system is utilized. The survey played a critical role in identifying key requirements, challenges, and user preferences, which directly influenced the development process.

\subsection{Survey}
 The secure exchange of medical records between professional and patient history tracking is the main motivation for filling the gap in EHR adoption.
In addition to the growing use of AI in the healthcare field, that offers solutions such as medical data history summarization, X-ray classification, and predictive analytics to assist doctors in diagnosing and treating patients more efficiently.
However, the level of awareness, acceptance and concerns of AI-integrated EHR systems among Egyptian doctors and patients is still unclear.
\subsubsection{Survey Design and Objectives}
A survey was conducted among Egyptian doctors and patients to explore the awareness and perceptions of EHRs and AI-assisted medical systems. This survey was conducted for the following objectives:
\begin{enumerate}
    \item To evaluate the level of knowledge about EHR systems among doctors and patients.
    \item To evaluate their opinions on the importance and usefulness of EHRs in improving medical data management.
    \item To understand their concerns and perceived challenges regarding the implementation of EHRs.
    \item To gather feedback on the potential role of AI in enhancing EHR functionalities and the level of trust they have for AI in the medical field.
\end{enumerate}
\subsubsection*{Target Audience}
The survey targeted two primary groups:
\begin{enumerate}
    \item Health professionals: Practicing in different affiliations (e.g., public and private hospitals or clinics in Egypt), with varying levels of experience and specialization (e.g., doctors, dentists and pharmacists)
    \item Patients: Individuals who visit doctors frequently or have long-term medical conditions requiring regular check-ups.
\end{enumerate}

\subsubsection{Survey Structure}
The survey was conducted in Egyptian Arabic and was divided into two sections for covering questions for the two groups. The survey was conducted using Google Forms and distributed online to reach a diverse group of respondents.  The survey link was shared through professional networks and social media platforms to gather a well-rounded dataset. Responses were collected over a specified period of two months to ensure an adequate sample size for meaningful analysis.

\subsubsection*{Medical Professionals' questions}
The question topics for this group:
\begin{itemize}
    \item General awareness of EHR systems: (e.g. "Have you heard of or used an EHR system before?")
    \item Concerns about HER systems: What are your primary concerns regarding EHR systems? (e.g. data security, system reliability, user interface, learning curve, technical barriers, time required for data entry)
    \item Perceived usefulness of EHRs in improving healthcare: (e.g. "Do you believe that current dependency on physical records needs to be changed?")
    \item Awareness and perception of AI in EHRs (e.g. "Would you trust AI-assisted summarization of medical records in helping doctors?")
\end{itemize}
 \subsubsection*{Patients’ questions}
  The question topics for the patients:
 \begin{itemize}
     \item General awareness of EHR systems: (e.g. "Have you heard of an EHR system before?")
     \item Their preference for medical affiliation (for example, “Usually, do you go to a private hospital, a public hospital, or a clinic?”)?
     \item How Important EHR to be implemented (e.g. "Do you believe an EHR system would improve healthcare accessibility ?")
     \item Medical Record-Keeping Habits (e.g. "How do you currently manage your medical records (e.g., prescriptions, test results, diagnoses)? Do you keep physical copies, digital records, or rely on memory?")
     \item Concerns and trust in EHRs:(e.g. "What concerns do you have about using an EHR system? (e.g., privacy, data security, access control, technical reliability")).
     \item Functionalities they wish get implemented in the system
\end{itemize}
\subsection{User Interface Design Overview}
According to the survey responses and the simple flat design standards, the Electronic Health Record (EHR) website is designed to erase the potential concerns health professionals and patients have.  
The website is designed with three distinct user views, each associated with a specific role within the healthcare system.
\begin{enumerate}
    \item Administrator’s View – Manages access control and doctors’ information in the hospital, Patient’s data addition requests and adding new patients to the system.
    \item Patient’s View – Enables patients to access their medical records, request examinations and data addition, schedule a visit and track their health status.
    \item Doctor’s View – Allows authorized doctors to review patient history, conduct examinations, and upload medical data.
\end{enumerate}
\subsubsection{User Authentication and Registration Process}
The user authentication process in our project is based on a strong yet simple system that promotes ease of access for everyone. \\
Administrators have sole access through predefined credentials, allowing for monitored supervision of the system. Patients and Doctors, in contrast, can register an account by providing personal information. National ID is a must for patient registration, acting as a key identifier in case of an emergency and supporting quick and specific information retrieval.
\subsubsection{Administrator’s View}
\begin{figure}[h!]
    \centering
    \begin{mdframed}[linecolor=blue, linewidth=0.25pt]
        \includegraphics[width=\linewidth]{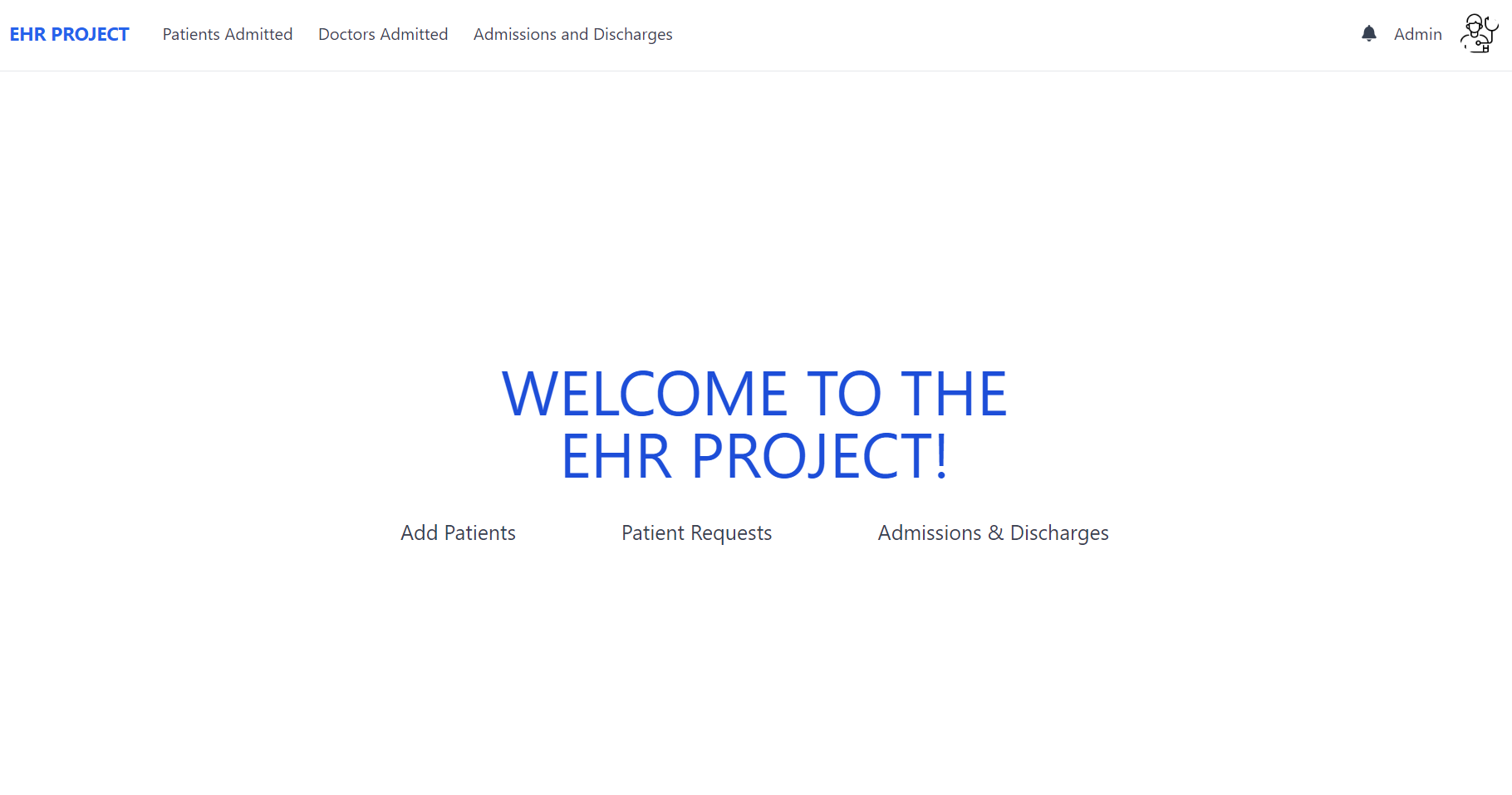}
    \end{mdframed}
    \caption{Administrator's View}
    \label{admin}
\end{figure}

The administrator plays a crucial role in managing hospital workflows, ensuring secure access to patient data. The administrator view is shown in figure \ref{admin}
The features provided to the administrator include:
\begin{itemize}
    \item \textbf{Add Patients}: If a patient did a hospital visit to ask for an examination admission without being registered on the system, it enables administrators to register new patients into the system.
    \item \textbf{Making Admissions}: Admins grant doctors access to patient records upon examination requests done by patient or hospital visits.
    \item \textbf{Making Discharges}: After a consultation or follow-up is completed, the administrator revokes the doctor’s access to the patient's records to maintain data privacy.
    \item Patient Data Requests: Manages data addition requests sent by patients, forwarding them to the respective doctors for verification and approval.
    \item Requests: Displays pending data addition requests and new examination requests.
    \item \textbf{Doctors and Admissions}: Provides a list of doctors with their assigned patient admissions, ensuring streamlined management of medical personnel.
    \item \textbf{Patients and admissions}: Provides a list of patients with their assigned doctor admissions, ensuring streamlined management of medical personnel.
\end{itemize}

\subsubsection{Patient’s View}
\begin{figure}[h!]
    \centering
    \begin{mdframed}[linecolor=blue, linewidth=0.25pt]
        \includegraphics[width=\linewidth]{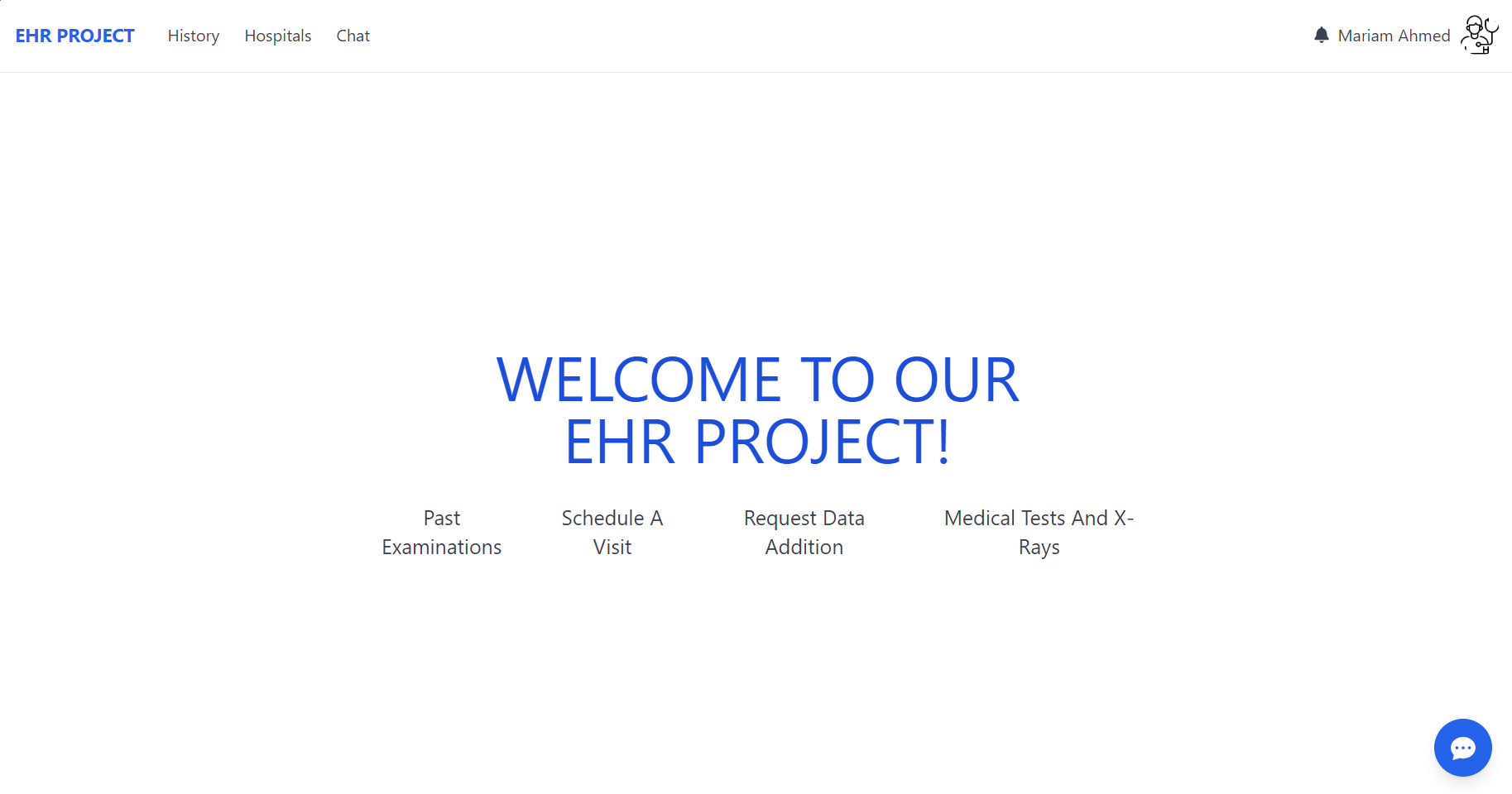}
    \end{mdframed}
    \caption{Patient's View}
    \label{patient}
\end{figure}

Patients interact with the system primarily to access their medical history, request medical addition, and access x-rays and lab test results. The patient dashboard is shown in figure \ref{patient}
The patient benefit from the following functionalities:

\begin{itemize}
    \item \textbf{Examinations}: Displays a list of past medical visits, including: Examination ID, Examination type, Date of visit, Doctor’s name, 
    \item \textbf{Details of Each Examination}: Offers a detailed view of each 
    examination’s findings:
    Complaints, Diagnosis, Symptoms, Treatments and Medications: Prescribed treatments and drug dosages, Doctor’s Notes.
    \item \textbf{Request Data Addition}: If there is data that is not stored by the system, Patients can request the addition of medical records (e.g., previous diagnoses and surgeries, test results) by submitting: Data Type (e.g., test results, prescriptions, reports), Issuance Date, Supporting Documents.
    \item Medical Tests and X-ray Results: Allows patients to view their uploaded lab results and radiology scans reports.
    \item Request Examination: Patients can request a new examination appointment through the system.
    \item \textbf{History}: Provides access to a patient's complete medical 
    record, including:
    \begin{itemize}
    \item Allergies
    \item Chronic conditions
    \item Past surgeries
    \item Current medications
    \item Immunizations
    \end{itemize}
    \item \textbf{Hospitals}: Displays a list of hospitals within the EHR network where the patient can seek treatment.
\end{itemize}

\subsubsection{Doctor’s View}
\begin{figure}[h!]
    \centering
    \begin{mdframed}[linecolor=blue, linewidth=0.25pt]
        \includegraphics[width=\linewidth]{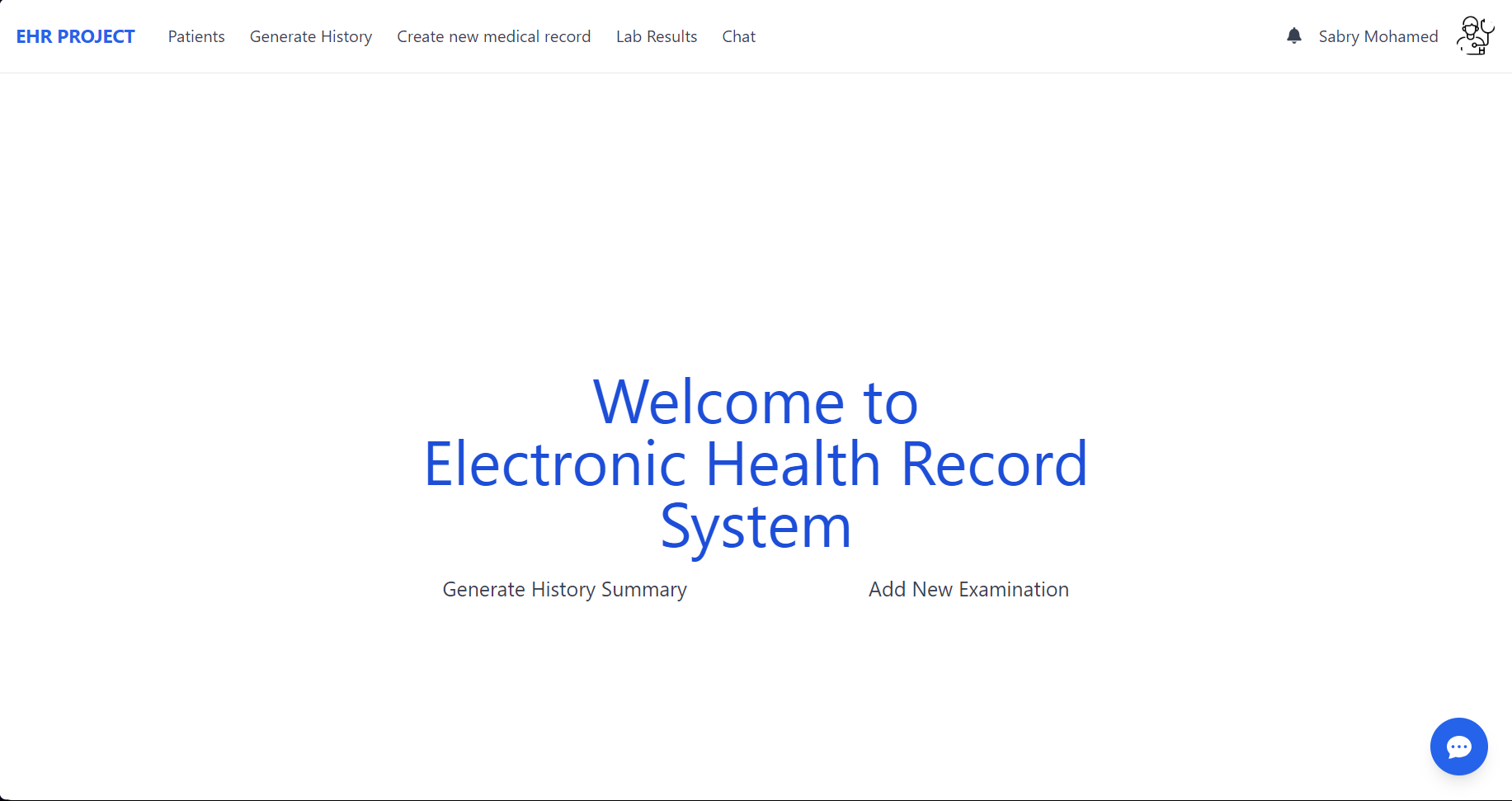}
    \end{mdframed}
    \caption{Doctor's View}
    \label{doctor}
\end{figure}

The doctor’s view is designed to provide a set of tools for patient care, using a AI advanced capabilities, and simplifying medical work for them. Doctor dashboard is shown in figure \ref{doctor}.
Doctors can benefit from the following functionalities:
\begin{itemize}
    \item \textbf{AI-powered Patient History Summarization:} Doctors can use an AI model to generate concise summaries of a patient’s medical background. All relevant information such as allergies, long-term medical conditions, past operations, current drugs, and vaccinations is included in such a summary. The AI summary enables quick comprehension of a patient’s background.
    \item \textbf{Create New Examination for Patients:} Doctors can log new exams for patients, specifying the kind of examination (e.g., routine examination, follow-up, emergency), date of examination, early symptoms, and any additional relevant information or guidance. All relevant information is documented for future use in such a feature.
    \item \textbf{Create New Medical Records for Patients:} Doctors can update and add new medical information for patients, such as basic information like name, age, contact information, and current drugs. With such a feature, patient files become updated and accurate at all times.
    \item \textbf{AI-powered X-ray Analysis:} Doctors can upload X-ray images for analysis by the AI model. The model used provides diagnostic feedback, such as identifying issues (e.g., fractures, tumors, infection) and recommending additional tests or interventions. This feature enhances the accuracy with which doctors can make diagnoses.
    \item \textbf{Review and Update Medical Records:} Doctors can view a patient’s medical record, including previous exams, laboratory tests, X-ray images, and received treatments. They can review and update such information when necessary to ensure everything is correct.

\subsubsection{Chatbot Integration}
Chatbot functionality was integrated in patient and doctor views. It serves as a virtual assistant that provides an immediate support and medical advice. Available at any time, the chatbot enables effective and efficient communication and guidance that enhances the efficiency for both medical professionals and patients.
\end{itemize}

\subsection{Agile-Driven Development}
The development of the Electronic Health Record (EHR) system was guided by a structured, iterative approach to address the technical and domain-specific challenges of Egypt’s healthcare ecosystem. This section outlines the software development lifecycle (SDLC) for delivering a scalable, secure, and AI-augmented EHR system.
\subsection{Architecture Overview: Microservices Approach}
This section outlines the microservices-based architecture, the rationale behind choosing microservices over a monolithic design, and the database and caching strategies employed to ensure performance and compliance with healthcare standards.

\textbf{Monolithic architecture: } In a monolithic architecture, all components of an application—such as the user interface, business logic, and database access—are tightly coupled and deployed as a single unit. While this approach simplifies initial development and deployment, it poses significant challenges for large-scale systems like EHRs:

\begin{itemize}
    \item Scalability: Scaling a monolithic application requires replicating the entire system, even if only one component faces increased demand.

\item Maintainability: Changes to one component can  affect others accidentally, which increases the risk of bugs and downtime.

\item Technology Lock-in: Monolithic systems are often built using a single technology stack that limits flexibility in adopting new tools or frameworks.\cite{micro1}
\end{itemize}

\textbf{Microservices Architecture:}
In contrast, a microservices architecture decomposes an application into smaller, loosely coupled services, each responsible for a specific business function. These services communicate via APIs and can be developed, deployed, and scaled independently. Key advantages include: \cite{micro2}

\begin{itemize}
    
    \item Scalability: Individual services can be scaled based on demand, optimizing resource utilization.
    
    \item Maintainability: Teams can work on different services simultaneously, reducing development bottlenecks.
    
    \item Technology Flexibility: Each service can use the most appropriate technology stack for its function.
    
    \item Fault Isolation: Failures in one service do not cascade to others, improving system resilience.
    
\end{itemize}

\begin{figure}[h]
    \centering
    \includegraphics[width=1\linewidth]{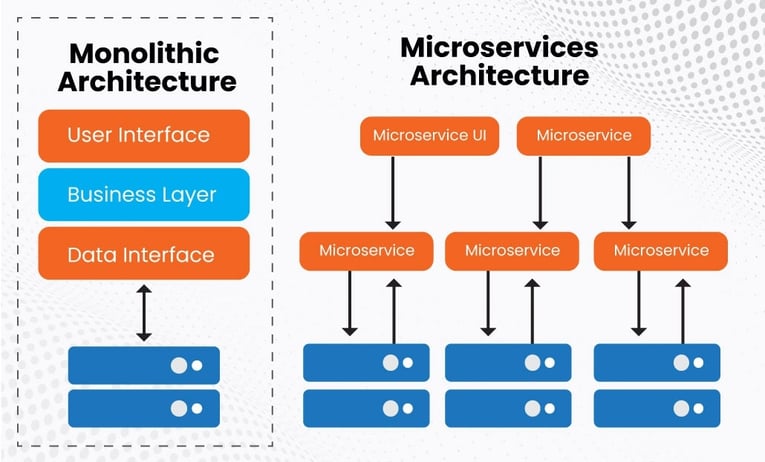}
    \caption{Microservices vs Monolithic}
    \label{microservices}
\end{figure}

The difference between the two architectures can be depicted in figure \ref{microservices}.

\subsubsection*{User Service}
\begin{figure}[h]
    \centering
    \includegraphics[width=\columnwidth]{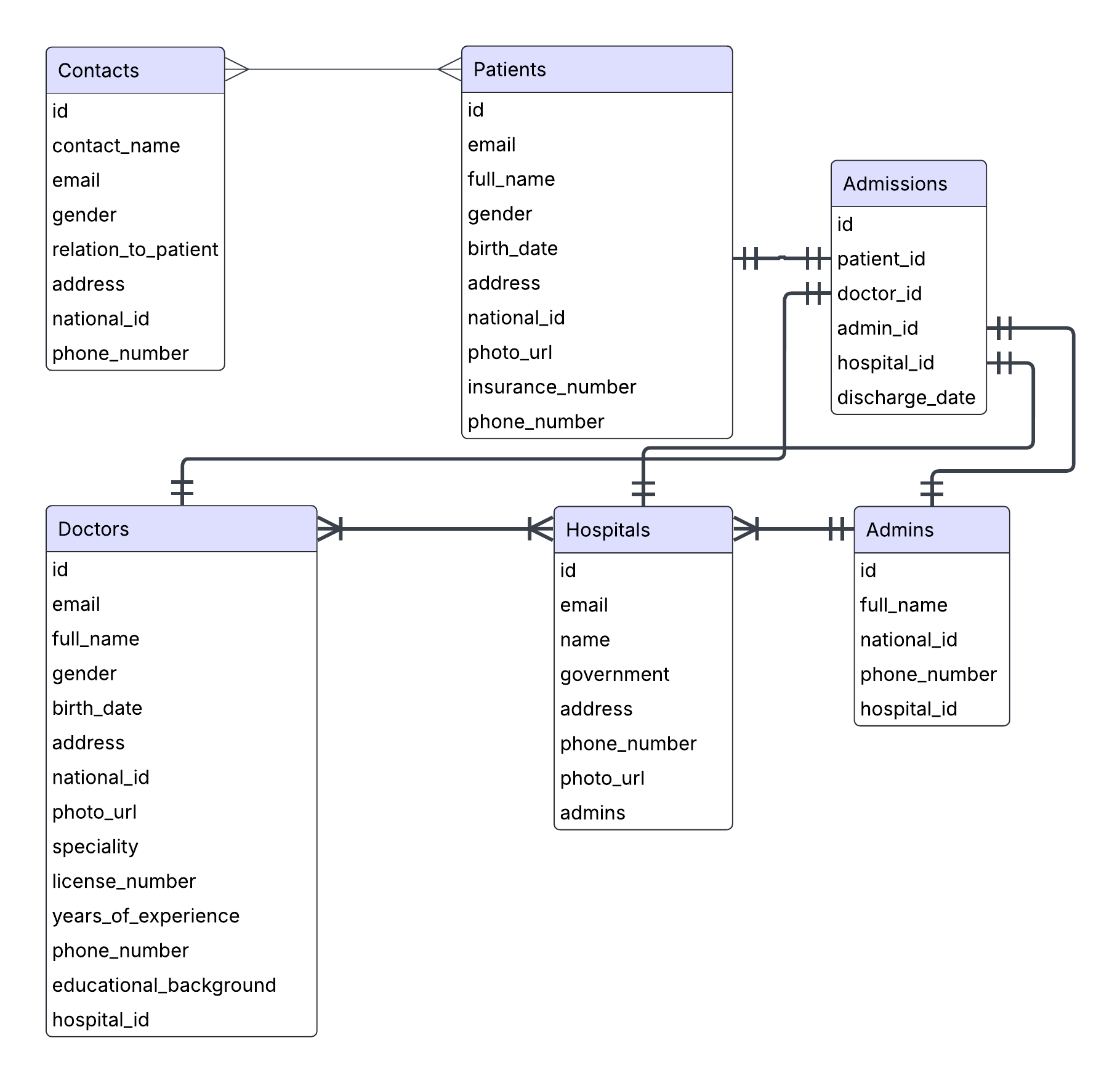}
    \caption{User-service ERD}
    \label{user-service_ERD}
\end{figure}

The user service is responsible for handling all aspects related to user profiles, authentication, and authorization. This is achieved through an implementation that relies on Auth0 to handle authentication and authorization. 
There are 3 user roles within the system: patients, doctors, and admins. All the relationships between them are illustrated in the Entity Relationship Diagram (ERD) diagram in figure \ref{user-service_ERD}.
\vspace{0.5cm}
\begin{enumerate}
    \item Patients
        \begin{itemize}
            \item Patients can register to create new accounts on the system and get authenticated when they log in with their email and passwords.
            \item They can list all hospitals within the system and get information about them.
            \item Patients can assign emergency contacts to them through a many-to-many relationship with the contacts table, as shown in figure \ref{user-service_ERD}.
            \item Patients can request admission to the hospital from the hospital admins (receptionists), who will assign them to a doctor with a one-to-one relationship stored in the admissions table, as shown in figure \ref{user-service_ERD}.
        \end{itemize}
        
    \item Doctors
        \begin{itemize}
            \item Doctors can be assigned to multiple hospitals within the system.
            \item Doctors are assigned to patients through an admin.
        \end{itemize}

    \item Admins
        \begin{itemize}
            \item They are responsible for managing the system; they can register patients and doctors and assign them to each other. They can view and track all admissions within the hospitals.
        \end{itemize}
\end{enumerate}

In addition, the user service creates entries in the Logs table to record significant events, such as access attempts, and assignments to provide an audit trail for security purposes and system monitoring.

\subsubsection*{Patient Service}
\begin{figure}[h]
    \centering
    \includegraphics[width=1\linewidth]{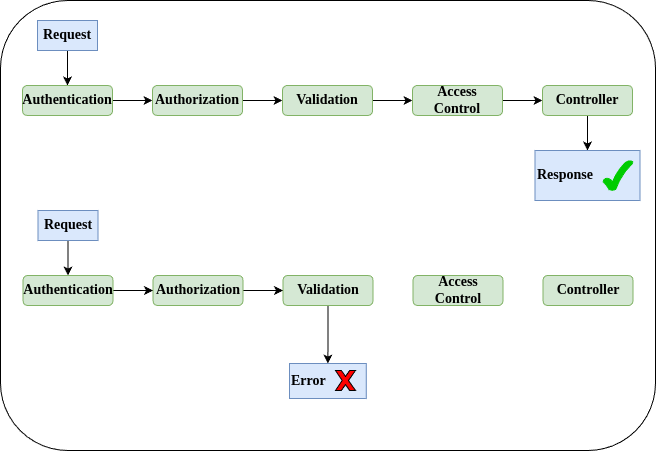}
    \caption{Request Cycle}
    \label{request-cycle}
\end{figure}

The patient service manages the storage and handling of all patient medical data. This includes medical records, conditions, examination details, allergies, surgeries, medications, and lifestyles of the patient. The service provides a standardized API for accessing and manipulating patient records efficiently. All incoming requests are passed through 5 sequential security layers, as illustrated in figure \ref{request-cycle}. A failure at any security layer will throw an error, and will not pass the request to the subsequent layer.

\begin{enumerate}
    \item Authentication: This is used to verify the identity of the user making the request to the patient service. The user is authenticated with Auth0 using an access token (JWT), that is included in the request. The patient service ensures that this token is valid, has not expired, and is issued by a trusted party.

        \begin{lstlisting}[language=JavaScript, caption=Authentication, label=Authentication, ]
export const authAccessToken = auth({
  audience: auth0_audience,
  issuerBaseURL: `https://${auth0_domain}`,
});

export const authenticate = (req, res, next) => {
  authAccessToken(req, res, err => {
    if (err) {
      return res.status(401).send('Unauthorized');
    }
    return next();
  });
};
        \end{lstlisting}
    \item \textbf{Authorization Layer:} This layer determines what actions an authenticated user is permitted to perform. For example, only doctors are allowed to create a new visit document for the patient. Patients attempting this action will receive a \texttt{ForbiddenError}. This is accomplished by determining certain permissions in Auth0 and assigning them to each user role. For instance, only doctors are allowed to create visit documents for patients because they are the only users who have \texttt{"createVisit"} permission.

            \begin{lstlisting}[language=JavaScript, caption=Authorization, label=Authorization]
export const authorizeUser = requiredRole => (req, res, next) => {
  try {
    // Extract the user's role from the token payload
    const { permissions } = req.auth.payload;

    // Check if the user has the required role
    if (!permissions.includes(requiredRole)) {
      throw new ForbiddenError('Forbidden: Insufficient permissions');
    }

    // If authorized, proceed to the next middleware or controller
    next();
  } catch (error) {
    console.error('Authorization error:', error);
    return sendError(res, error);
  }
};
        \end{lstlisting}

    \item \textbf{Validation:} The validation layer ensures that the incoming data is well-structured and meets the required format to be stored within the database. This is accomplished by using Joi, a powerful schema description language and data validator for JavaScript. If an incoming request contains an invalid data format, a \texttt{ValidationError} will be thrown. 

    \item \textbf{Access Control:} This is an additional security layer that ensures that only relevant users can access and manipulate the data they are permitted to handle. While authorization ensures that only doctors can access a patient's data, access control restricts access only to the assigned doctor to this patient. This improves the system's security over sensitive information.
\end{enumerate}

\begin{figure}[h]
    \centering
    \includegraphics[width=\linewidth]{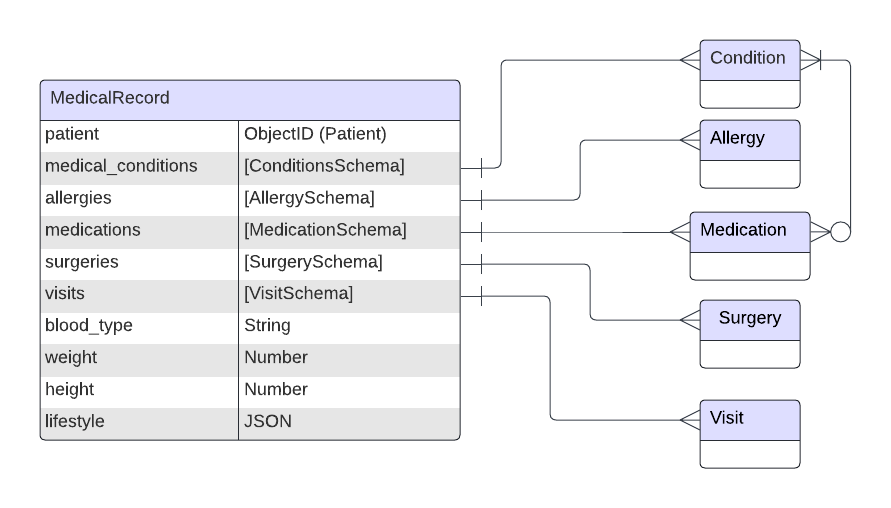}
    \caption{Abstract patient-service ERD}
    \label{abstract_erd_patient}
\end{figure}

Each patient within the system should have only a single medical record document in the database (one-to-one). Within this document, all his medical history is stored and linked to other collections. As shown in figure \ref{abstract_erd_patient}, the medical record document contains information about the patient’s allergies, medical conditions, medication, surgeries, and visits. The assigned doctors are permitted to create new documents related to a patient's care, such as new visit records. Once created, these documents are automatically linked to the patient's medical record, ensuring a precise and up-to-date view of medical history.

\subsubsection*{AI Service}

The AI service mainly manages the generation of medical summaries based on the patient's medical history, the chatbot functionality, the AI-driven medical reporting and recommendation, and the chest x-ray functionality. These functionalities are important in accelerating the interaction between doctors and patients, resulting in more time to give care to the patients. The service uses the Llama3 OpenBioLLM-70B model for the generation tasks and vit-xray-pneumonia classification for the x-ray classification. The service is further explained in the Artificial Intelligence Integration section.

\subsection{Data modeling and Database Design:}
The EHR system employs a polyglot persistence strategy \cite{martin_fowler}, leveraging multiple databases to optimize data storage and retrieval for different types of information. This section details the data modeling approach, the rationale behind choosing MongoDB for patient records, and the schema design for key entities.
\subsubsection{Data Modeling Approach}
During the requirements gathering phase, interviews with Egyptian clinicians revealed that modeling a comprehensive medical record containing every detail about a patient for example, modeling and storing 
all neural or cardiology physical exams, would be inefficient and impractical. Instead, doctors emphasized the need for a general yet critical dataset that is universally relevant across departments, practices, and healthcare providers.
Based on this feedback, the team adopted an aggregate data model \cite{martin_fowler}, where each patient’s medical record is composed of key entities that encapsulate the most critical information:
\begin{enumerate}
    \item \textbf{Medical Conditions}: Chronic (e.g., diabetes, hypertension) and non-chronic (e.g., infections, injuries).

\item \textbf{Medications}: Current and past prescriptions, including dosage and frequency.

\item \textbf{Allergies}: Drug, food, and environmental allergies.

\item \textbf{Surgeries}: Past surgical procedures and outcomes.

\item \textbf{Lifestyle}: Habits such as smoking, alcohol consumption, and exercise.

\item \textbf{Hospital Visits/Exams}: Summaries of clinical encounters, including diagnoses, treatments, and follow-up recommendations.
\end{enumerate}
As evident from Figure \ref{abstract_erd_patient}, each entity is modeled independently and stored in its own collection, enabling granular data management. A medical record document aggregates these entities into a single, unified view for efficient access and analysis. This approach ensures that the system captures the most relevant patient data while maintaining flexibility for future extensions. Dealing in aggregates makes it much easier for databases to handle operating on a cluster since the aggregate makes a natural unit for replication and sharding, which MongoDB supports to ensure the system can handle large volumes of patient records.

\subsubsection{Caching}
Another thing to keep in mind is that medical systems often operate on huge amounts of data daily. A large system like EHR would be data intensive on the databases alone and would affect the performance of the system. In order to lighten the burden, a caching layer was implemented using Redis. An in-memory key-value database. It's mainly used to optimize performance for frequently accessed data like:

\begin{itemize}
    \item \textbf{Active User Sessions:} Reduces database load by storing session data in memory.
    \item \textbf{Frequent Queries:} Caches commonly accessed patient data (e.g., recent lab results).
    \item \textbf{AI Outputs:} Stores LLaMA 70B responses to improve response times for similar queries.
\end{itemize}

The system utilizes both relational and non-relational databases to manage different data types efficiently.
\begin{enumerate}
    \item User-service (PostgreSQL): \\
     PostgreSQL is an open-source object-relational database management system, which performs better in handling structured and well-defined data. It’s known for its support of ACID (Atomicity, Consistency, Isolation, and durability) properties, which are important in providing integrity in transactional operations. Further, PostgreSQL handles complex relationships between tables effectively through foreign keys. All these unique attributes of PostgreSQL make it align perfectly with the requirements of the \textbf{user-service} to manage complex relationships between users in the system.
     
    \item Patient-service and AI-Service (MongoDB): \\
    MongoDB is a document-oriented NoSQL database program. It’s classified as a NoSQL since it uses JSON-like documents with optional schemas. This flexible storage mechanism makes it highly adaptable to handle unstructured data. The patient-service deals with complex patient data often containing nested information and varying data types. Therefore, the \textbf{patient-service} and \textbf{AI-service} can greatly benefit from this flexibility of MongoDB schema.
\end{enumerate}

\subsection{Technology Stack}
\subsubsection{Frontend Development}
React with Vite:
  \begin{itemize}
      \item React was chosen as the core front-end framework due to its component-based architecture, which allows for reusability and modular development, making it easier to maintain and scale the project.
      \item Vite was selected over traditional build tools like Webpack because of its fast development server, hot module replacement (HMR) for quick updates, and optimized production builds.
  \end{itemize}
TypeScript (TSX):
  \begin{itemize}
      \item TypeScript was used instead of JavaScript to enhance type safety as it reduces runtime errors by catching issues at compile time.
      \item It improves code readability and maintainability, making collaboration among developers easier.
      \item Provides better IntelliSense and autocompletion, increasing developer productivity.
  \end{itemize}
HTML \& CSS:
  \begin{itemize}
      \item HTML was used for structuring the application's UI components.
      \item CSS was used for styling and layout for a responsive and visually appealing interface.
      \item Tailwind CSS was integrated for faster styling and to maintain consistency in UI components.
  \end{itemize}
\subsubsection{Backend Development}
The backend of the \textbf{user-service} and \textbf{patient-service} are developed using Express.js, a lightweight and commonly used node framework. Node and Express.js were chosen as they provide some key features that will enhance the development and user experience of the proposed system, such as:

\begin{itemize}
    \item Optimized throughput and scalability.
    \item Speed of development.
    \item Ease of Implementation.
\end{itemize}

Express provides mechanisms to write handlers for HTTP requests at different URL paths and build server-side applications with support for various modules.
In addition, the AI service is developed using Python and FastAPI for efficient integration with machine learning libraries and suitability in building APIs.

\subsection{Security and Compliance}
This section outlines the multi-layered security approach adopted to protect patient data, prevent unauthorized access, and ensure compliance with relevant standards. The security measures span authentication, authorization, validation, role-based access control (RBAC), audit logging, and encryption, providing a robust defense against potential threats.

\subsubsection{Authentication and Authorization}
There are different strategies to implement authentication and authorization:
\subsubsection*{Authentication and Authorization on each service}
one approach is to implement authentication in every component of the system individually and enforce it on each entry point. This would be beneficial if different security policies and authentication strategies need to be present in the same system but, this approach requires meticulous planning as it's harder to maintain and monitor. Additionally, this introduces duplication between services and each microservice would then depend on user authentication data which it might not own and even poses a great risk if this data is shared with multiple services.
\subsubsection*{Global Authentication and Authorization Service}
In this strategy, a dedicated microservice handles the authentication and authorization. And each service must pass the request to the authentication service before processing it. However, this also has several downsides one of them being that authorization checks are a business concern governed by business rules which are related to each microservice. Therefore, authorization should not be handled with another entity that is not related to the business.\\
After studying several authentication strategies, the best solution for our use case is to utilize a third-party identity provider Auth0 to manage authentication and authorization.
\textbf{Authentication with Auth0}: The system leverages Auth0 as its identity provider to manage user authentication securely. Auth0 was chosen for its compliance with industry standards (e.g., OAuth 2.0, OpenID Connect) and its ability to handle complex authentication scenarios, such as multi-factor authentication (MFA) and social logins.

Workflow:
\begin{enumerate}
    \item Users log in to our system handled by the user management and authentication microservice, which communicates with Auth0's authentication server using its SDK to authenticate the user.
    \item Upon successful authentication, Auth0 issues a JSON Web Token (JWT) containing user claims (roles, user information, permissions) shown in Listing \ref{lst:decoded-jwt}.
    \item Users interact with the system providing the JWT which is validated by the target microservice using Auth0’s, ensuring its integrity and authenticity before processing.
\end{enumerate}

\begin{lstlisting}[language=json, caption=Decoded JWT Example in JSON format, label={lst:decoded-jwt}, xleftmargin=0.5cm]
{
  "user_roles": [
    "doctor"
  ],
  "iss": "https://ehr-ai.eu.auth0.com/",
  "sub": "auth0|67902deee3e34f9f52b9dd15",
  "aud": [
    "https://gateway-ehr-api.com",
    "https://ehr-ai.eu.auth0.com/userinfo"
  ],
  "iat": 1738340572,
  "exp": 1738426972,
  "scope": "openid profile email",
  "gty": "password",
  "azp": "23o3Jc4C3JpaDya46Yy2jSBeLzJX1oN5",
  "permissions": [
    "createAllergy",
    "createCondition",
    "createMedication",
    "createRecord",
    "createSurgery",
    "createVisit",
    "deleteAllergy",
    "deleteCondition",
    "deleteMedication",
    "deleteSurgery",
    "deleteVisit",
    "getRecord",
    "getVisit",
    "updateAllergy",
    "updateCondition",
    "updateMedication",
    "updateSurgery",
    "updateVisit"
  ]
}
\end{lstlisting}

\paragraph{Role-Based Access Control (RBAC)}
The system implements RBAC to restrict access to functionalities based on user roles. For example, doctors can view and edit medical records for their patients, while patients can only view their own records. Admins manage user accounts and system settings but cannot access medical data. Access control ensures data consistency and prevents misuse. Notably, patients are restricted from editing their records to prevent self-diagnosis and ensure that all medical data is entered by qualified practitioners.\\
Additionally, another form of access control is enforced through an admission system. For Example, the doctor role has permissions to access and edit medical records if the patient is admitted to the medical facility, that the doctor has these privileges. After the person finishes his visit / exam, the admin discharges the patient and the doctors are no longer allowed to view this patient's records. This complies with Egypt's Data Protection Law.

\begin{figure}[h]
    \centering
    \includegraphics[width=\linewidth]{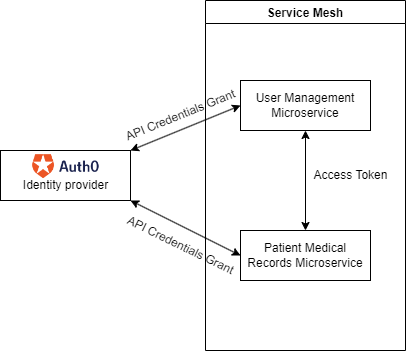}
    \caption{Machine-to-machine Authentication}
    \label{m2m-auth}
\end{figure}

\paragraph{Microservice Authentication}
Microservices communicate using machine-to-machine (M2M) authentication, a feature provided by Auth0. In this model, each microservice is registered as a client in Auth0 and is issued a client ID and client secret. When one microservice needs to call another, it requests an access token from Auth0 using its credentials. The target microservice validates the token using Auth0’s public keys before processing the request. A diagram of the process can be seen in Figure \ref{m2m-auth}

This approach ensures that only authorized microservices can communicate with each other, preventing unauthorized access to sensitive data. For example, the AI Integration microservice must present a valid access token to interact with the Patient Records microservice. This token contains scopes that define the specific actions the AI service is allowed to perform, such as reading patient data or generating reports. 
\subsubsection{Validation Across Layers}
The validation layer ensures that the incoming data is well-structured and meets the required format to be stored within the database. This is accomplished by using Joi, a powerful schema description language and data validator for JavaScript. If an incoming request contains an invalid data format, a ValidationError will be thrown as seen in Figure \ref{request-cycle}. 

\subsubsection{Audit logs}
To detect and prevent malicious activities, the system maintains audit logs that record all user actions. These logs include timestamps, user IDs, actions performed, and additional context, such as changes made to patient records. Audit logs are stored in MongoDB, enabling efficient querying and analysis. They are immutable and retained for a minimum of five years to comply with Egypt’s Data Protection Law. This logging mechanism ensures accountability and provides a reliable trail for investigating potential security incidents.

An example of the audit logs can be seen in the code snippet \ref{lst:audit-logs}.

\begin{lstlisting}[language=json, caption=Patient Medical History Record  Example in JSON format, label={lst:audit-logs}, xleftmargin=0.5cm]]

{   
    "_id": 67981ea4e7f5329a8682971a
    collection_name: "medical_records"
    document_id: 6798175b4c21fc495ff6d06e
    action: "VIEW"
    doctor_id: "uth0|67902deee3e34f9f52b9dd15"
    ip_address: "48.224.122.4"
    user_agent: "Nginx V1.2"
    reason: "Medical Record Viewed"
    access_type: "Regular"
    status: "Success"
    createdAt: 2025-01-28T00:02:44.911+00:00
    updatedAt: 2025-01-28T00:02:44.911+00:00
    version: 0
}

\end{lstlisting}

\subsubsection{Encryption and Secure Communication}
The system employs encryption mechanisms to protect data at rest and in transit. Sensitive data stored in PostgreSQL and MongoDB is encrypted using AES-256, while Transparent Data Encryption (TDE) is enabled for MongoDB to encrypt entire databases. For data in transit, all communication between clients and the API gateway is encrypted using TLS 1.3, ensuring that sensitive information cannot be intercepted. Additionally, mutual TLS (mTLS) is used for service-to-service communication within the Kubernetes cluster, providing an additional layer of security for internal traffic.

\subsubsection{Rate Limiting and DoS Protection}
To prevent denial-of-service (DoS) attacks, the system implements rate limiting on API endpoints. A global rate limit of 1,000 requests per minute per IP address is enforced, along with a user-specific limit of 100 requests per minute per authenticated user. These limits are enforced at the API gateway using NGINX and Redis for tracking request counts. This approach ensures that system resources are protected from abuse while maintaining availability for legitimate users.

\subsubsection{Compliance with Egyptian Regulations}
The system is designed to comply with Egypt’s Personal Data Protection Law (Law No. 151 of 2020), which mandates data minimization, access control, audit trails, and data breach notification. Only necessary data is collected and stored, and strict RBAC policies limit access to sensitive information. Immutable audit logs are maintained for accountability, and mechanisms are in place to detect and report data breaches within 72 hours. These measures ensure that the system adheres to regulatory requirements and builds trust among users.

\subsection{Artificial Intelligence Integration}

\subsubsection{Role of AI in the system}
\begin{itemize}
    \item Chatbot Functionality: The chatbot is a conversational interface for users to interact with the EHR system. It runs mainly on the Llama3 OpenBioLLM model. It provides real-time responses to natural language queries. This ensures optimal accessibility for both doctors and patients.\\
    For Doctors, the chatbot allows healthcare providers to retrieve patient information efficiently, query specific details, and access AI-generated insights during consultations. For example, a doctor can ask, “What are the patient’s recent lab results?” or “Summarize the history of diabetes treatment for this patient.” The system instantly provides a concise and actionable response. This helps the doctors in the decision-making process.
    For Patients: Patients can use the chatbot to view their medical history, receive updates on upcoming appointments, and access personalized health tips.
    \item Medical History Summarization: The EHR system uses the Llama3 OpenBioLLM to analyze and synthesize patient data into concise, actionable summaries. This feature significantly reduces doctors' cognitive load by offering quick overviews of complex medical histories. This facilitates the decision-making process.
    Examples of summarization include:
    \begin{enumerate}
        \item Condensing a patient’s medication history into a single paragraph to highlight essential treatments and outcomes.
        \item Identifying and summarizing critical trends, such as increasing cholesterol levels or recurring diagnoses, to assist in proactive healthcare management.
    \end{enumerate}
    \item Medical Reporting and Recommendations Service for Last Visit: The proposed EHR system automatically generates a medical report according to the last visit. This medical report will include the patient’s information, visit details, diagnosis, medications, vitals, and AI recommendations for medical recommendations, predictive insights, abnormal values, and lifestyle recommendations. 
    \item X-Ray Pneumonia Classification: Multiple models have been implemented for disease detection to aid with the classification of the patient. The ViT X-Ray Pneumonia Classification model is used to classify the patient’s X-ray for having the Pneumonia disease or not. Therefore, the doctor will have time to just review the classification result and more care-time for the patient.
\end{itemize}
\subsubsection{AI Model Selection and Overview}
\subsubsection*{Llama3-OpenBioLLM-70B}
Meta-Llama-3-70B-Instruct, Llama-3 for short, is a collection of foundational language models ranging from 7B to 65B that was presented in 2023 as part of Meta AI. Llama3.1 demonstrated superior performance compared to many mainstream LLMs such as GPT-3 and PaLM across multiple benchmarks, showcasing its capability to handle complex biomedical tasks.\cite{grattafiori2024llama3herdmodels}
Llama3-OpenBioLLM-70B was selected as the foundational AI model for this system due to its specialization in biomedical contexts and superior performance. The model’s architecture and pretraining make it uniquely suited for tasks such as summarization and predictive analysis in healthcare.\cite{OpenBioLLMs}

Biomedical Specialization: OpenBioLLM-70B is pretrained from Llama-3-70B-Instruct on a diverse set of large-scale, high-quality biomedical datasets to accustom the unique language and knowledge requirements of the medical field. 

\begin{figure}[h]
    \centering
    \includegraphics[width=1\linewidth]{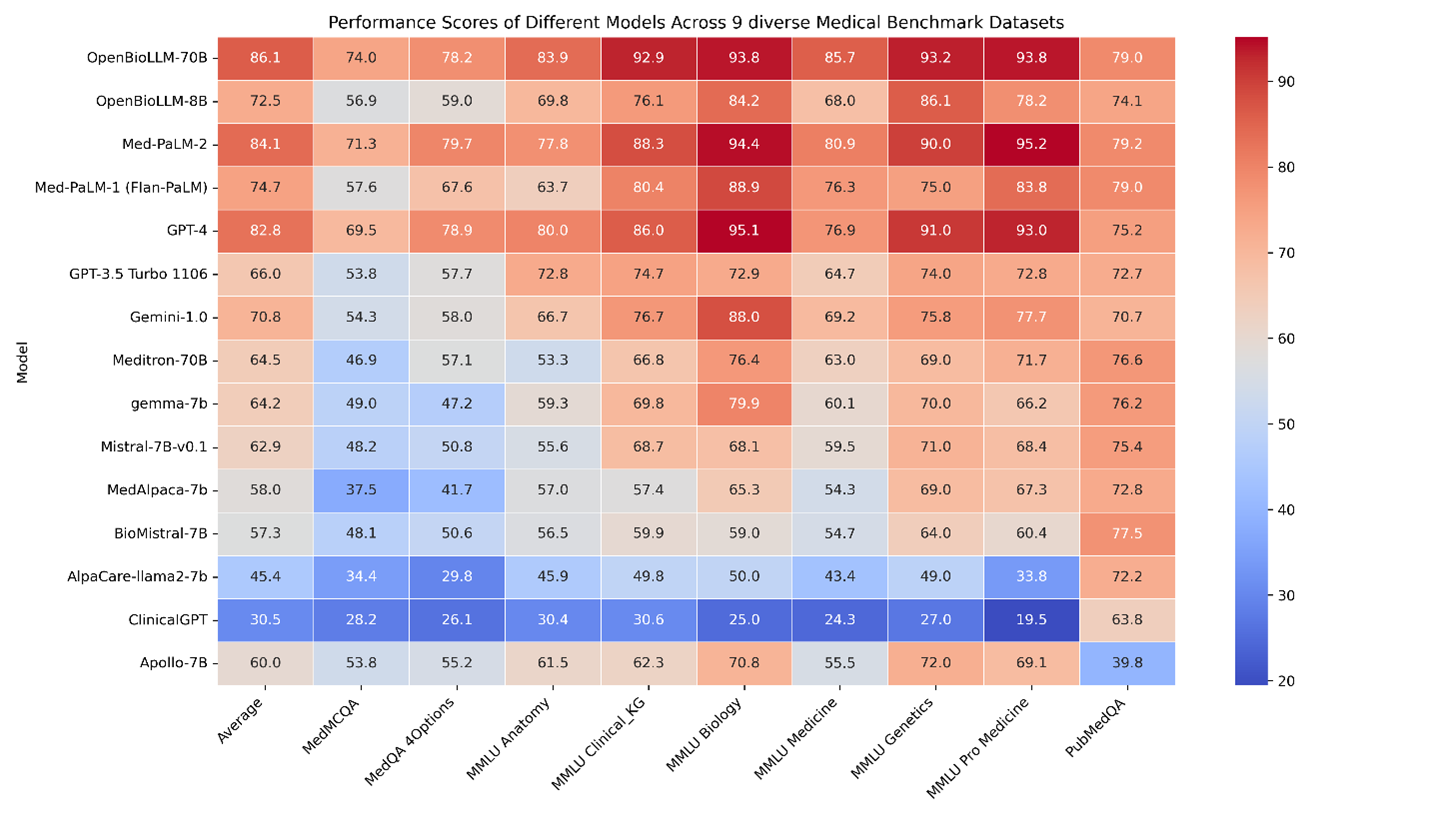}
    \caption{Performance Scores of Different Biomedical Models Across Multiple Benchmarks}
    \label{performance}
\end{figure}

Superior Performance: With 70 billion parameters, Llama3-OpenBioLLM-70B excels in handling biomedical tasks with unmatched precision. It has consistently outperformed larger proprietary and open-source models, including GPT-4, Gemini, Meditron-70B, Med-PaLM-1, and Med-PaLM-2, on key biomedical benchmarks as shown in figure \ref{performance}.

 Model Architecture: Llama3-OpenBioLLM-70B is based on a transformer architecture, which uses a decoder-only mechanism designed for autoregressive tasks such as text generation and summarization. Unlike encoder-decoder architectures typically used for translation tasks, decoder-only models focus solely on generating meaningful outputs by processing the input sequence and predicting subsequent tokens.
 
These capabilities make Llama3-OpenBioLLM-70B an integral part of the EHR system, addressing the dual challenges of information overload and decision-making complexity in healthcare environments.

\subsubsection*{ViT-Xray-Pneumonia-Classification}

 Unlike traditional convolutional neural networks (CNNs), ViTs process images as a sequence of patches This allows them to capture long-range dependencies and spatial relationships more effectively. This makes ViTs  suitable for medical imaging applications, where subtle patterns in X-ray scans can indicate underlying conditions such as pneumonia. \cite{vit1}\cite{vit2}

For the EHR system, the ViT x-ray pneumonia classification model is fine-tuned on Google’s ViT-BasePatch16-224 using the National Institute of Health (NIH) Chest X-rays dataset.\cite{xray-dataset} It achieves a loss of 0.0868 and accuracy of accuracy: 0.9742. The model will enhance diagnostic efficiency while assisting radiologists in medical decision-making.

\subsubsection{AI Workflows and Integrations}

\subsubsection*{Summarization Workflow}

The summarization workflow begins with retreiving the pateint's medical history: conditions, medications, visits, allergies, and life-style from MongoDB, the patient database. This data is fed into the Llama3 OpenBioLLM model along with a system role for the summarization task.

\begin{enumerate}
    \item Defining the System Role and Prompt: The AI model is guided by a system role and the user prompt. These roles defines the model's purpose and scope to ensure medical relevance and response quality.

    \item Fetching and Preprocessing Patient Data: When a doctor requests a summary via the front-end, the FastAPI backend retrieves the patient’s data from MongoDB based on the provided patient ID. The raw data is stored in JSON format and then converted into a structured text format that aligns with the AI’s processing requirements.

    \item Model Processing and Summarization: The processed patient history is sent to the Llama3 OpenBioLLM model, which extracts key medical insights and generates a structured summary. This summary is then displayed on the doctor’s dashboard for review.

\end{enumerate}

\subsubsection*{Chatbot Workflow}

The chatbot workflow begins with front-end interface, where the user could start a new chat. The back-end and Llama model handles the interaction between users and the EHR system by natural language processing to understand queries and generate precise responses. The process includes the following detailed steps and views:

\begin{itemize}
\item Scenario 1: When a doctor initiates a new conversation, they provide their unique identifier (doctor\_id) and the initial query (user\_input) to the /chat/initiate/ endpoint. The system generates a unique conversation\_id to track the session and assigns a predefined system\_role to the chatbot to contextualize its role as a medical assistant. The Llama3-OpenBioLLM-70B model processes the input and generates an AI-driven response (bot\_reply). The system then saves the conversation details, including the conversation\_id, query, and response, into the database. Finally, the system returns the conversation\_id and the AI-generated bot\_reply to the doctor, as shown in the sequence diagram in Figure \ref{scenario1}.
           \begin{figure}[h]
                \centering
                \includegraphics[width=1\linewidth]{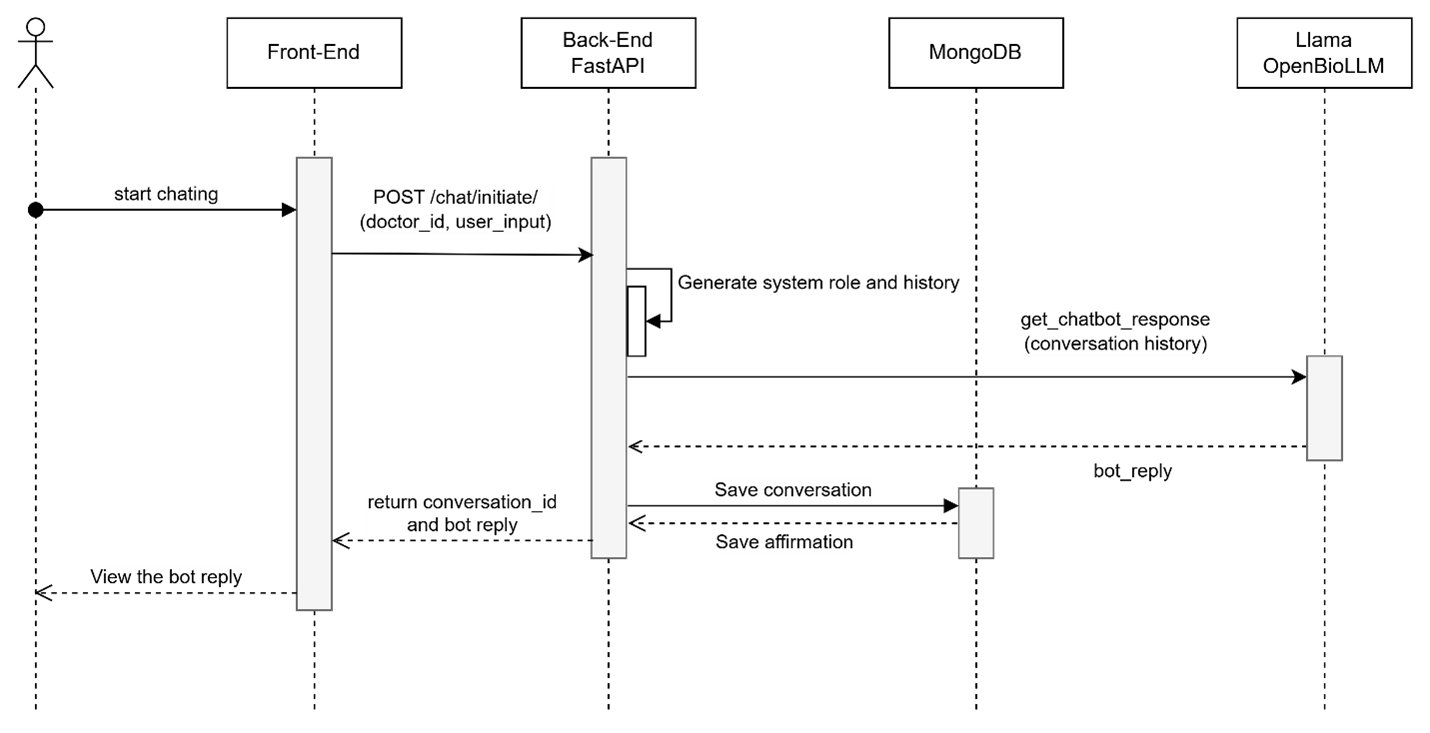}
                \caption{Sequence Diagram of Scenario 1: Start a New Chat Session}
                \label{scenario1}
            \end{figure}
    \item In Scenario 2, the workflow focuses on continuing an existing conversation, as illustrated in the sequence diagram of Figure \ref{scenario2}. The process begins when the doctor selects an existing conversation ID from the /chats/ endpoint. If a complete conversation history is required, it can be retrieved using /chat/conversation id. The doctor then inputs a new query, which is submitted to the /chat/continue/ endpoint. The system appends this new query to the existing conversation log. The updated context, including the latest input, is sent to the Llama3-OpenBioLLM-70B model for processing, generating a new AI-driven response. Once the response is generated, the system updates the conversation log in the database with this latest interaction. Finally, the system returns the AI-generated reply to the doctor, seamlessly continuing the conversation, as shown in the sequence diagram in Figure \ref{scenario2}.
     \begin{figure}[h]
                \centering
                \includegraphics[width=1\linewidth]{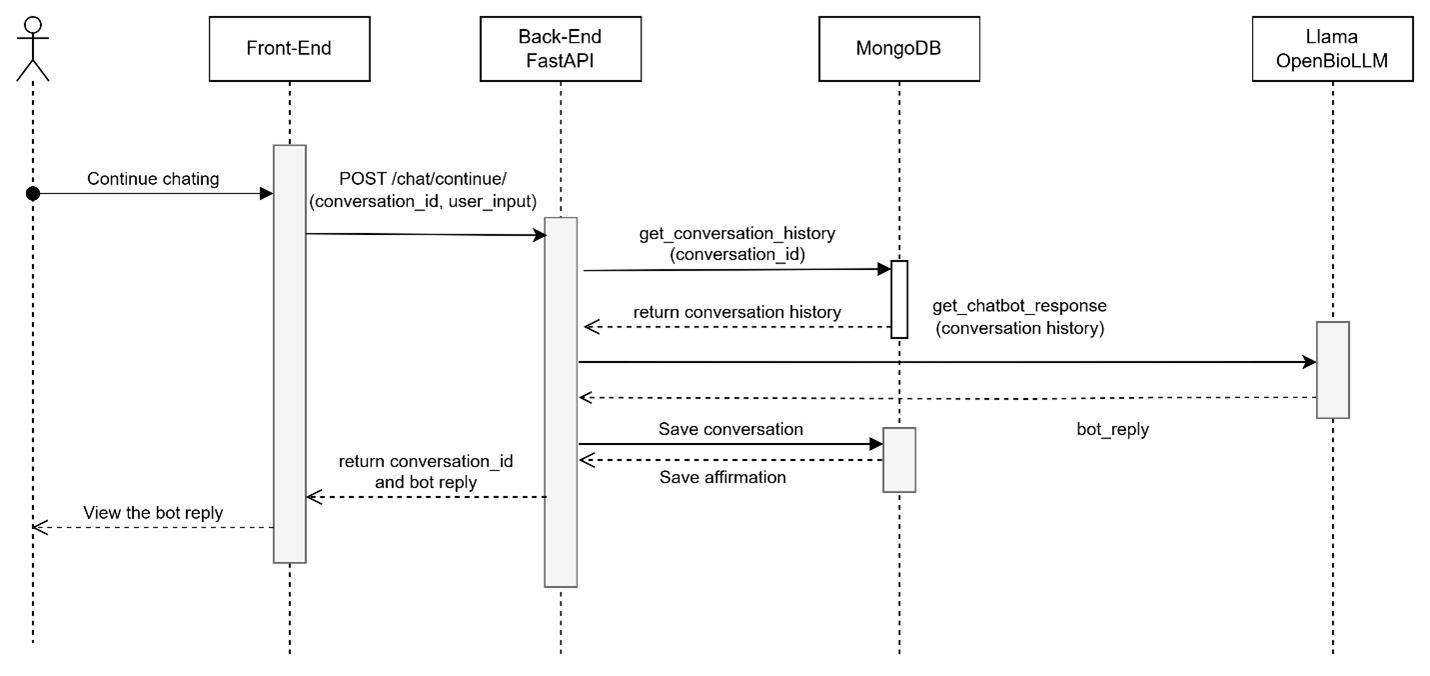}
                \caption{Sequence Diagram of Scenario 2: Continue an Existing Conversation}
                \label{scenario2}
            \end{figure}

    \item Integrated Workflow:
        \begin{enumerate}
            \item Fetch All Conversations: Doctors can view a list of all previous interactions via /chats/, allowing them to decide whether to continue an existing conversation or initiate a new one.
            \item Retrieve Full History: Detailed conversation logs can be accessed using /chat/{conversation\_id}.
            \item Initiate or Continue: Depending on the scenario, either /chat/initiate/ or /chat/continue/ is called, with the AI model seamlessly integrating to provide contextually accurate responses.
            \item Save Updates: All interactions are logged in the chatbot\_history collection of the patient\_service database (MongoDB) for future reference as examplified in figure \ref{chat-history-example}.

            \begin{figure}[h]
                \centering
                \includegraphics[width=1\linewidth]{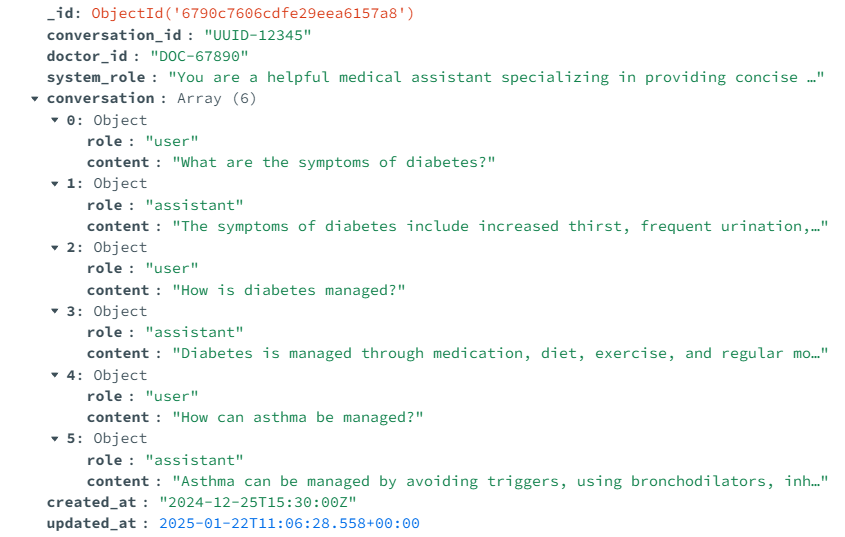}
                \caption{Chat History Example in the Database}
                \label{chat-history-example}
            \end{figure}
        \end{enumerate}
\end{itemize}

\subsubsection*{Auto-generated Medical Report Workflow}

The Auto-generated Medical Report feature uses the patient data and a specified visit to automatically generate structured medical reports. This reduces the time spent on manual documentation and allows healthcare providers to focus on patient care. The report includes key details such as patient information, visit summaries, diagnoses, prescribed medications, vitals, and AI-driven recommendations. The workflow is
\begin{enumerate}
    \item Triggering Report Generation
    \begin{itemize}
        \item When a patient’s consultation is completed, the system automatically generates a medical report.
        \item Alternatively, a doctor can request a report manually through the interface.
    \end{itemize}

    \item Fetching and Preprocessing Patient Data
    \begin{itemize}
        \item The FastAPI backend retrieves the latest patient visit details from MongoDB using the patient ID.
        \item Relevant structured and unstructured data (e.g., vitals, lab results, diagnoses, medications, clinical notes) are extracted.
        \item The data is transformed from JSON format into a structured, natural-language summary to be fed into the AI model.
    \end{itemize}

    \item AI Processing and Summarization
    \begin{itemize}
        \item The Llama3-OpenBioLLM-70B model receives the preprocessed data along with a structured prompt defining the system’s role in report generation.
        \item The model analyzes the data and generates a concise, human-readable medical report.
        \item AI-driven recommendations, including predictive insights and lifestyle suggestions, are embedded in the report.
    \end{itemize}

    \item Report Formatting and Output Generation
    \begin{itemize}
        \item The AI-generated summary is formatted into a standardized medical report template.
        \item Sections include:
        \begin{itemize}
            \item \textbf{Patient Information}: Demographics and basic health profile.
            \item \textbf{Visit Summary}: Date, reason for visit, and key findings.
            \item \textbf{Diagnosis \& Treatment}: Physician-assigned diagnoses and prescribed medications.
            \item \textbf{Vitals \& Lab Results}: Key physiological measurements and flagged abnormalities.
            \item \textbf{AI Recommendations}: Predictive risk factors, abnormal value alerts, and lifestyle guidance.
        \end{itemize}
    \end{itemize}

    \item Saving and Accessing the Report
    \begin{itemize}
        \item The generated report is saved in Azure Storage and the AI recommendations are saved in MongoDB for future reference.
        \item It is made available in the doctor’s dashboard, where it can be reviewed, modified, or exported as a PDF.
    \end{itemize}
\end{enumerate}

\subsubsection*{X-ray Classification Workflow}

The X-ray Classification Module in the proposed EHR system automates pneumonia detection in chest X-rays using Vision Transformer (ViT) models. This workflow ensures a seamless process from image upload to AI-assisted diagnosis.
\begin{enumerate}
    \item X-ray Image Upload and Preprocessing
    \begin{itemize}
        \item A doctor or radiologist uploads a DICOM or JPG/PNG image via the EHR system.
        \item The backend converts and resizes the image to 224×224 pixels for model compatibility.
    \end{itemize}

    \item Feature Extraction \& Model Inference
    \begin{itemize}
        \item The ViT Xray Pneumonia Classification model extracts essential visual features from the image.
        \item The model analyzes these features and predicts whether the X-ray suggests Pneumonia or Normal.
    \end{itemize}

    \item Prediction Generation
    \begin{itemize}
        \item The model outputs a classification label along with a confidence score.
        \item Example output:
        \begin{itemize}
            \item Classification: Pneumonia
            \item Confidence Score: 92\%
        \end{itemize}
    \end{itemize}

    \item Integration with Medical Records
    \begin{itemize}
        \item The AI-generated diagnosis is appended to the patient’s EHR record for review.
        \item The radiologist can confirm, modify, or override the AI’s decision.
    \end{itemize}

    \item Displaying Results
    \begin{itemize}
        \item The system presents:
        \begin{itemize}
            \item The X-ray image alongside the AI’s classification.
            \item A doctor’s override option to ensure human validation.
        \end{itemize}
    \end{itemize}

    \item Saving and Reporting
    \begin{itemize}
        \item The AI decision is stored in MongoDB under the patient’s diagnostic history.
        \item The X-ray image is stored in Azure Storage service.
    \end{itemize}

\end{enumerate}

\subsection{System Deployment Architecture}
This section discuss the technical implementation of our deployment strategy, encompassing containerization, deployment automation, container orchestration, and cloud infrastructure utilization.

\subsubsection{Container-based Architecture}
Our system implements containerization using Docker as the foundation of the deployment strategy. Each microservice and its dependencies are encapsulated within isolated, portable units, ensuring consistency across different environments. The containerization architecture employs multi-stage Dockerfile configurations to optimize resource utilization and enhance security.
The container security framework includes non-root user execution and systematic exclusion of development artifacts through careful configuration. This approach ensures that each microservice maintains isolation while operating in a consistent environment, whether deployed locally for development or in production on cloud infrastructure.

\subsubsection{Deployment Automation}
\begin{figure}[h]
    \centering
    \includegraphics[width=\linewidth]{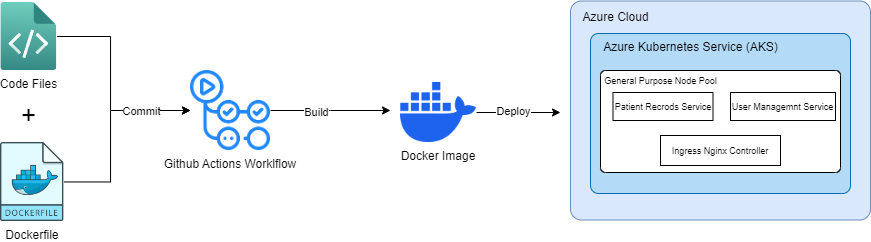}
    \caption{CI/CD pipeline flow for microservice deployment to AKS}
    \label{fig:cicd-pipeline}
\end{figure}

The deployment pipeline leverages GitHub Actions to automate the software delivery process. During the build phase, each microservice undergoes compilation and packaging into a Docker image. These images are versioned using Git commit hashes, ensuring complete traceability throughout the deployment lifecycle. The Azure Container Registry serves as a centralized repository for all container images, providing secure storage and efficient distribution.
Helm charts manage the deployment configuration, handling environment-specific parameters through values files. This approach enables consistent deployment across different environments while maintaining the flexibility to adapt to specific environmental requirements. The automation pipeline ensures that deployments are reproducible and maintains system integrity throughout the delivery process. A high-level diagram of CI/CD pipeline is depicted in Figure \ref{fig:cicd-pipeline}.

\subsubsection{Container Orchestration Implementation}
Kubernetes serves as the cornerstone of our container orchestration strategy, managing the complex interactions between microservices. The orchestration layer implements deployment configurations that ensure high availability through multiple replica deployments and carefully specified resource allocations. Service architecture within the cluster facilitates internal communication through ClusterIP services, while external access is managed via a LoadBalancer service for the API Gateway.
Traffic management is handled by an NGINX ingress controller, implementing path-based routing to direct requests to appropriate microservices. Sensitive configuration data and credentials are managed through ConfigMaps and Secrets, maintaining security while enabling flexible configuration management.

\subsubsection{Cloud Infrastructure Design}
Our deployment uses Azure Kubernetes Service (AKS) as the primary cloud infrastructure platform. The cluster architecture implements specialized node pools to accommodate varying workload requirements. The infrastructure implements horizontal pod autoscaling to dynamically adjust resource allocation based on demand. Integration with Azure Active Directory provides good access control, while network policies enforce secure inter-service communication. Resource management is optimized through custom node pool configurations that align with specific service requirements, ensuring efficient resource utilization while maintaining system performance.

\subsection{Performance Testing}
This section details the performance testing conducted on the application's user profile retrieval endpoint using Grafana k6. The goal was to assess the application's responsiveness and stability under increasing user load. The tests simulated a realistic user scenario, fetching user profile data using GET requests to the /api/user/profile endpoint.

\subsubsection{Test Setup}
The performance tests utilized k6, a modern load-testing tool, to simulate concurrent users accessing the application. The test scenario involved a staged approach:
\begin{enumerate}
    \item \textbf{Ramp-up:} Over the first minute, the number of virtual users (VUs) gradually increased to 50, simulating increasing user traffic.

    \item \textbf{Sustained Load:} The 50 VUs were sustained for two minutes to assess the application's performance under constant load.

    \item \textbf{Ramp-down:} The number of VUs gradually decreased to 0 over the final minute, simulating decreasing user traffic.

\end{enumerate}

The K6 test script used to assess the system performance can be seen below in listing \ref{k6}:

\begin{lstlisting}[language=JavaScript, caption=Grafana K6 test script, label=k6, xleftmargin=0.6cm]
import http from 'k6/http';
import { check, sleep } from 'k6';

// Configuration
export let options = {
  stages: [
    { duration: '1m', target: 50 },  // Ramp up to 50 users over 1 minute
    { duration: '2m', target: 50 },  // Stay at 50 users for 2 minutes
    { duration: '1m', target: 0 },   // Ramp down to 0 users over 1 minute
  ],
  thresholds: {
    http_req_duration: ['p(95)<500'], // 95% of requests should complete within 500ms
    http_req_failed: ['rate<0.01'],   // Less than 1% of requests should fail
  },
};

// Test scenario
export default function () {
  // Set the headers with the access token
  const params = {
    headers: {
      'Authorization': `Bearer token`,
    },
  };

  // Fetch user's profile
  let profileUrl = 'http://4.255.82.40/api/user/profile';
  let profileRes = http.get(profileUrl, params);
  check(profileRes, {
    'Profile status is 200': (r) => r.status === 200,
  });

  // Simulate user think time
  sleep(1);
}

\end{lstlisting}

The test focused on the following key performance metrics:

\begin{itemize}
    \item \textbf{Response Time:} The time taken for the server to respond to a request, measured in milliseconds (ms). This metric directly reflects the user experience. The http-req-duration metric in k6 captures this.

    \item \textbf{Throughput:} The number of requests processed per second (requests/s). This metric indicates the system's capacity to handle concurrent requests. This can be calculated from http-reqs in the k6 results.
    
    \item \textbf{Error Rate:} The percentage of failed requests. This metric is critical for assessing the reliability and stability of the application under load. The k6 metric http-req-failed captures this directly.
\end{itemize}

\section{Results}
\subsection{Survey Responses and Insights }
\begin{figure}[h!]
                \centering
                \includegraphics[width=0.8\linewidth]{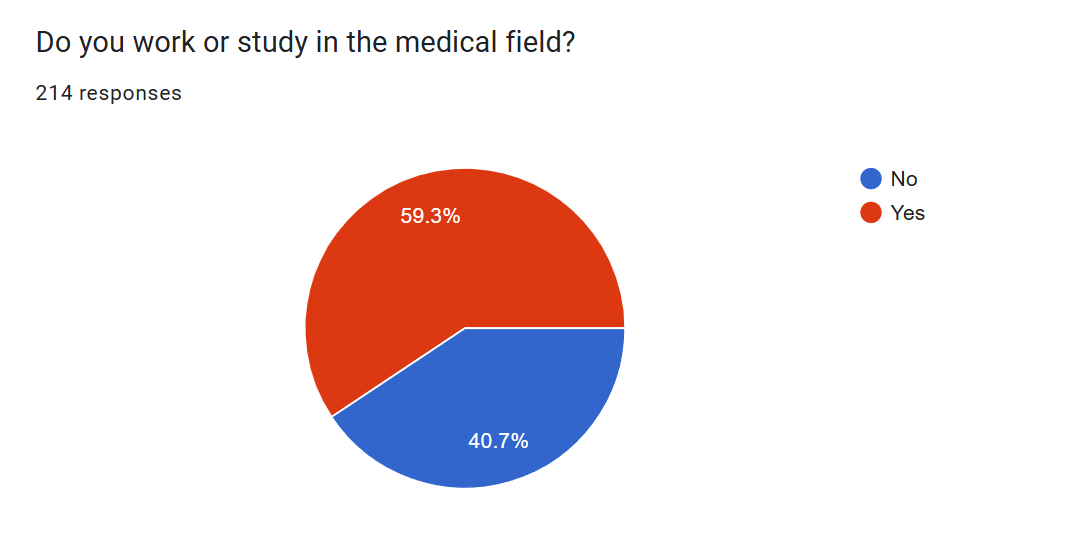}
                \caption{Ratio of Medical professionals to Patients.}
                \label{ratiodtop}
            \end{figure}

The survey was completed by a total of 220 people, consisting of 127 medical professionals and 93 patients and the results were translated to English, as shown in Figure \ref{ratiodtop}.
 \vspace{2cm}           
\subsubsection{Medical Professionals' Insights}
\begin{itemize}
    \item \textbf{EHR Awareness:} The majority of them had heard of EHR systems but had never used them, this percentage came with 63\%; on the other hand, 31.5\% had never heard of it and the rest are using it .
    \item \textbf{Urgency for Transition to EHRs:}
    91.3\% agreed that the current paper-based system should be replaced immediately with EHR systems as shown in fig\ref{update}.
    \begin{figure}[h]
                \centering
                \includegraphics[width=1\linewidth]
                {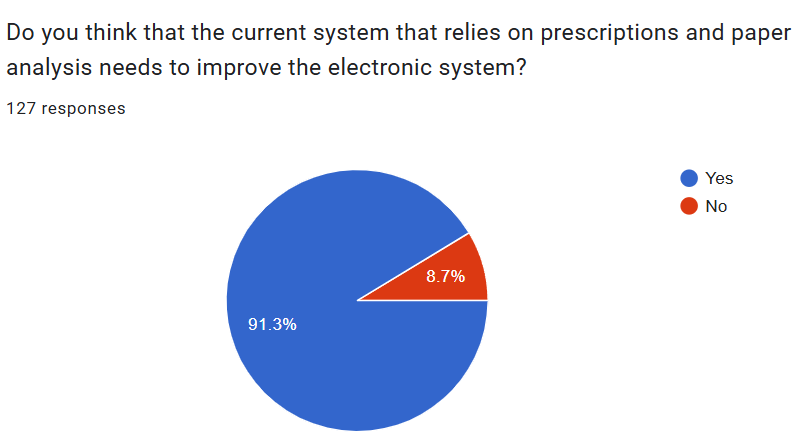}
                \caption{Urgency for EHR Adoption}
                \label{update}
            \end{figure}
     \item \textbf{Essential EHR Functionalities:}
    Most valued feature: Medical history record tracking for every patient and simple UI for fast usage as show in figure\ref{functiond}.
    \begin{figure}[h!]
        \hspace*{-0.7cm} 
        \includegraphics[width=1.3\linewidth, height=5cm]{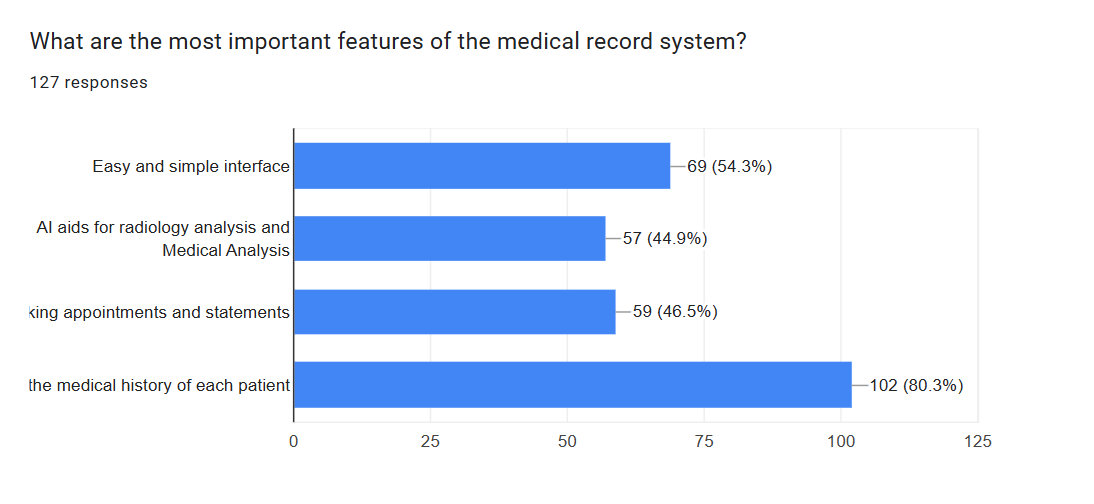}
        \caption{Functionalities needed for doctors.}
        \label{functiond}
        \end{figure}
        
    \item \textbf{Concerns About EHR Implementation:} 
    \begin{itemize}
        \item Main concern: System failure during emergencies
        \item Second major concern: Patients may struggle to use the system correctly
        \item Third major concern: No devices would be afforded in the affiliation for the system running and data entry
    \end{itemize}
    
    \item \textbf{AI Trust Issues:}
    66.1\% of medical professionals did not trust AI for summarizing medical test results and X-rays fig \ref{trust}.
    \begin{figure}[h]
                \hspace*{0.5cm} 
                \includegraphics[width=1\linewidth]
                {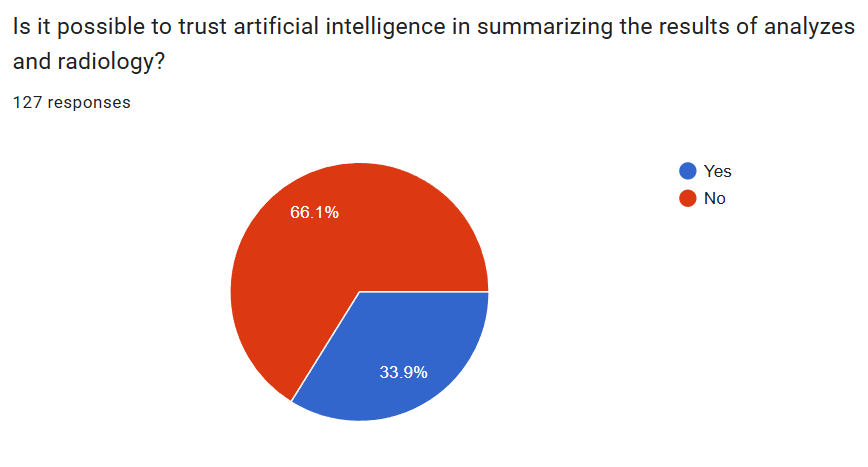}
                \caption{How much medical professionals trust AI in medical reports and X-rays}
                \label{trust}
            \end{figure}

    \item \textbf{Data Security Concerns:}
    \begin{enumerate}
        \item Primary concern: How patient data would be managed securely
        \item Preferred security measure: Encrypting patient data
        \item 8.7\% of medical professionals were strongly stressed about patients’ data security
    \end{enumerate}
    
    \item\textbf{Opinions on AI in EHRs:}
    Many professionals found AI useful for data organization, diagnosis suggestions, and treatment optimization. However, some doctors opposed AI intervention in decision-making, believing it should remain under direct medical supervision
\end{itemize}

\subsubsection{Patients' Insights}
\begin{itemize}
    \item \textbf{EHR Awareness and Willingness to Use:}
    \begin{enumerate}
        \item Majority of the patients hadn’t heard of the EHR systems before the survey
        \item Most saw value in digital records but had concerns about the technological barriers that would face users
        \begin{figure}[h]
                \centering
                \includegraphics[width=1\linewidth]
                {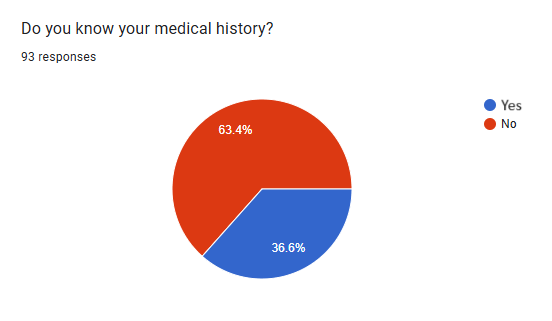}
                \caption{Patient's medical history knowledge}
                \label{medKnow}
            \end{figure}
        \item Medical Record-Keeping Habits 63.4\% of patients do not know their medical history as shown in fig \ref{medKnow}. In figure \ref{medcalrecords}, 31.2\% of patients get rid of their paper-based records after a while because they think they do not need them any more.
        \begin{figure}[h!]
                \centering
                \includegraphics[width=1\linewidth]
                {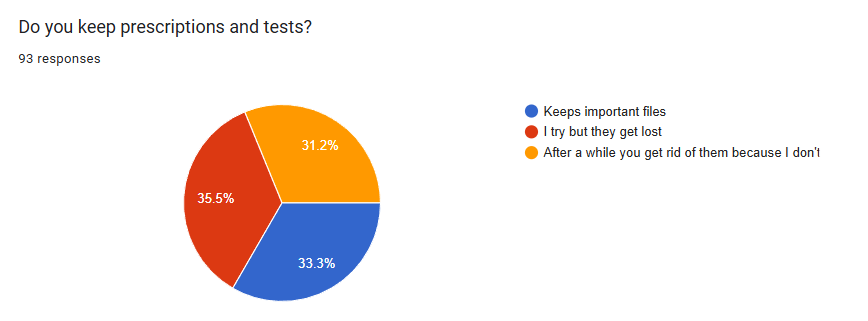}
                \caption{How much patients keep their medical records}
                \label{medcalrecords}
            \end{figure}
    \end{enumerate}
    
    \item \textbf{Concerns:}
    \begin{enumerate}
        \item Primary concern: 69.9\% agreed that it would not be utilized correctly by the specialist 
        \item Secondary concern: 50.5\% had some concerns on data privacy and system security and 21.5\% are not confident that such a system will keep their data private as shown in figure \ref{psecurity}    
        \begin{figure}[h]
                    \centering
                    \includegraphics[width=1\linewidth]
                    {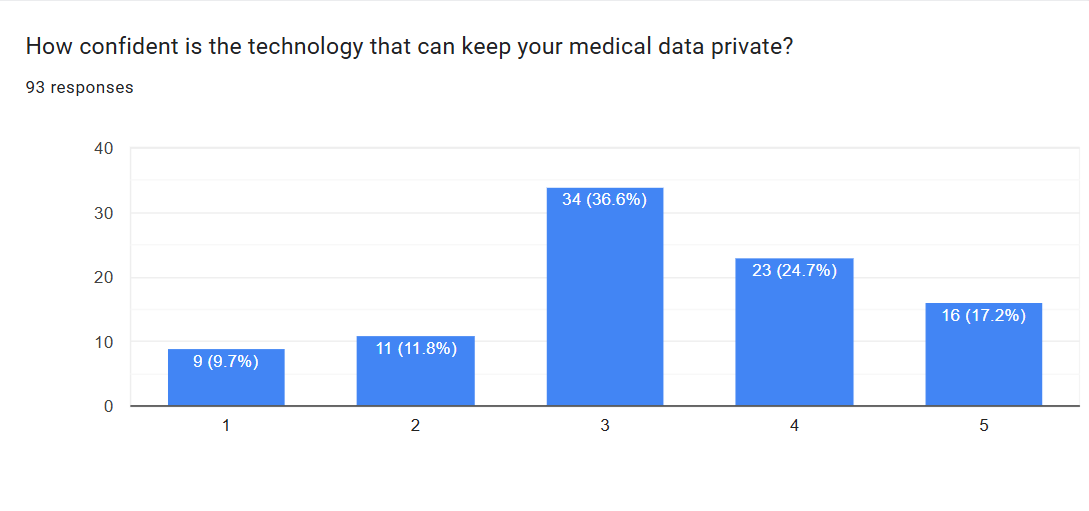}
                    \caption{Patients' trust Regarding Data Privacy and System Security}
                    \label{psecurity}
                \end{figure}
        \end{enumerate}
    \item \textbf{EHR Benefits Identified by Patients:}
    As shown in figure \ref{pfunction}.
    \begin{enumerate}
        \item 78.5\% chose that it will Ease access to medical history
        \item 74.2\% chose that it will reduce the risk of losing important documents
        \item 66.7\% agreed that it will help in scheduling visits in hospitals and clinics 
    \end{enumerate}
    \begin{figure}[h!]
                \centering
                \includegraphics[width=1\linewidth]
                {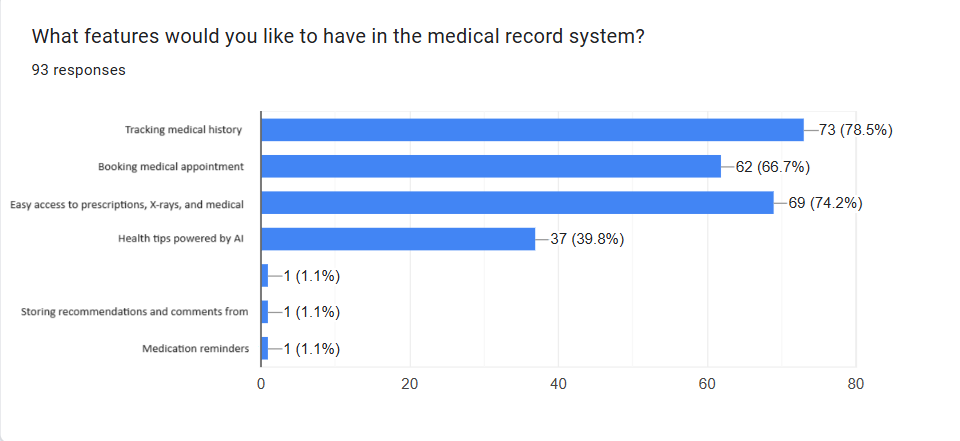}
                \caption{Functionalities patients need in the system}
                \label{pfunction}
            \end{figure}
    
\end{itemize}
\subsubsection{Notable Quotes from Medical Professionals}
\begin{itemize}
    \item "EHRs would help me easily access complete patient medical history, saving time and improving diagnosis and treatment decisions."
    \item "AI could be useful in analyzing patient history, prescriptions, and test results, identifying risk factors and potential drug interactions."
    \item "Having a structured and organized patient record system would improve diagnosis accuracy and treatment efficiency."
    \item "AI could assist in summarizing past test results, but final medical decisions should remain in human hands."
    \item "While AI may help with data structuring and organization, full reliance on it for diagnostics could be dangerous."
\end{itemize}

\subsection{Summarization Performance Evaluation for the AI}

Text summarization is an important task in the fields of Natural Language Processing (NLP). Its aim is to condense text into a more concise form. Thus, a quantitative metric is needed to evaluate its performance for the summarization task. In this section, ROUGE and BERTScore metrics are used for the evaluation of the AI-generated medical. The goal is to assess the model's efficiency in accurately summarizing patient medical histories while maintaining critical medical details. A thorough evaluation is essential to ensure that AI-generated summaries are both reliable and useful in clinical decision-making.

Summarization in medical applications requires both factual accuracy and conciseness. In healthcare, accurate summaries of the patient's medical history are important for diagnosis and treatment planning. Unlike traditional text summarization tasks, medical summarization requires a balance between retaining essential information and avoiding redundancy.

Since there is not medical record and medical history summary dataset available that fits the project's use case, a new dataset of medical history has been made. The dataset includes 150 medical histories as reference: 100 AI-generated medical histories and 50 human-written medical histories. The 100 AI-generated medical histories have been produced by ChatGPT and Llama3. This ensures variability in the input. Then, 150 medical history summarization by the Llama3 OpenBioLLM as the generated summaries. 

To quantify the performance of our AI-driven summarization model, metrics used as follows:
\begin{enumerate}
    \item ROUGE Metrics to evaluate lexical overlap between AI-generated summaries and ground-truth references, helping assess how much of the original content is retained.
    \item BERTScore to measure semantic similarity between the summaries, providing a deeper evaluation of meaning preservation rather than just word overlap.
\end{enumerate}

These evaluations help determine the system’s capability to generate clinically useful summaries for healthcare providers.

\subsubsection{ROUGE Score Analysis}

ROUGE (Recall-Oriented Understudy for Gisting Evaluation) is one of the most commonly used metrics for text summarization. It measures the overlap of words and phrases between the AI-generated summaries and the reference summaries. This guarantees the AI-generated content aligns with human-written summaries.\cite{rouge} The ROUGE family consists of several variants:
\begin{enumerate}
    \item \textbf{ROUGE-1}: Measures unigram (word-level) overlap, which helps determine how much of the original text is preserved.

    \item \textbf{ROUGE-2}: Measures bigram (phrase-level) overlap, providing insight into how well the AI captures sequential relationships between words.

     For any \(n\)-gram based ROUGE metric (with \(n=1\) for unigrams and \(n=2\) for bigrams), the recall, precision, and F1-score are defined as follows:

    \begin{align}
     ROUGE\text{-}n_{\text{recall}} = \frac{\sum\limits_{\text{match} \in n\text{-grams}} \text{Count}_{\text{match}}}{\sum\limits_{n\text{-grams} \in \text{reference}} \text{Count}}
\end{align}
     
    \begin{align}
    \hspace*{-0.5cm}
         ROUGE\text{-}n_{\text{precision}} = 
        \frac{\sum\limits_{\text{match} \in n\text{-grams}} \text{Count}_{\text{match}}}{\sum\limits_{n\text{-grams} \in \text{generated}} \text{Count}}
    \end{align}
    
    \begin{align}
        \hspace*{-0.7cm}
         ROUGE\text{-}n_{F1} = \frac{2\, ROUGE\text{-}n_{\text{precision}} \times ROUGE\text{-}n_{\text{recall}}}{ROUGE\text{-}n_{\text{precision}} + ROUGE\text{-}n_{\text{recall}}}
    \end{align}
    
    \item \textbf{ROUGE-L}: Measures the longest common subsequence (LCS), assessing fluency and coherence, which are important for readability.
    
    \begin{align}
        ROUGE\text{-}L_{\text{recall}} = \frac{LCS(\text{reference}, \text{generated})}{\text{length}(\text{reference})}
    \end{align}
    
    \begin{align}
        ROUGE\text{-}L_{\text{precision}} = \frac{LCS(\text{reference}, \text{generated})}{\text{length}(\text{generated})}
    \end{align}
    
    \begin{align}
        \hspace*{-0.8cm}
         ROUGE\text{-}L_{F1} = \frac{2 \times ROUGE\text{-}L_{\text{precision}} \times ROUGE\text{-}L_{\text{recall}}}{ROUGE\text{-}L_{\text{precision}} + ROUGE\text{-}L_{\text{recall}}}
    \end{align}
    
\end{enumerate}

\begin{figure}[h]
    \centering
    \includegraphics[width=1\linewidth]{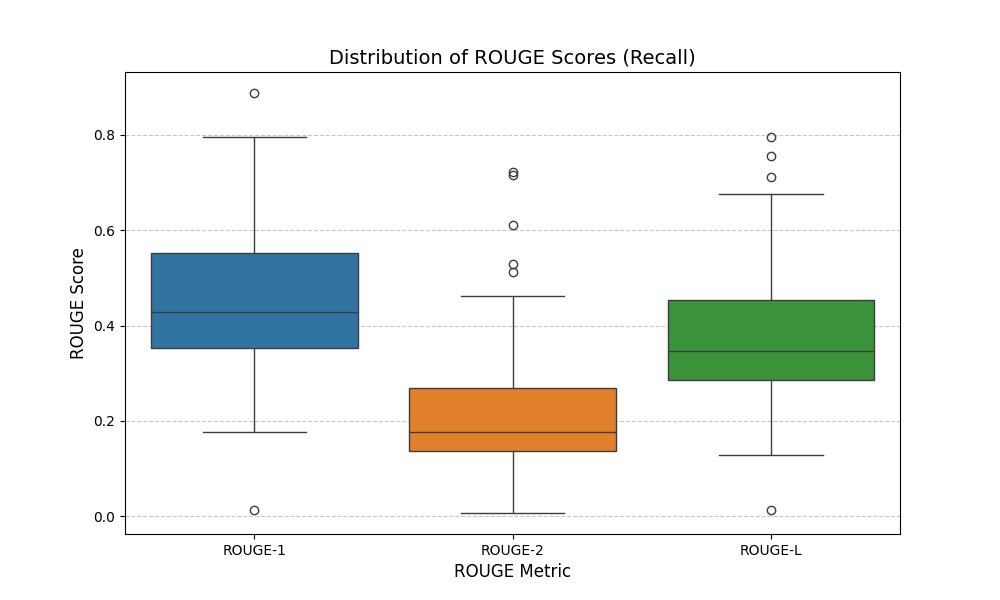}
    \caption{Distribution of ROUGE Scores for Recall}
    \label{recall}
\end{figure}

In ROUGE, \textbf{Recall} focuses on how much of the important information from the reference summary was captured in the AI-generated summary. A high recall means the AI didn’t miss out on key details. This is the most important metric for the medical summarization since it shows that all the words in the reference summary have been captured by the system summary. The results of the ROUGE recall of the ROUGE-1, ROUGE-2, ROUGE-L are shown in figure \ref{recall}.

\begin{figure}[h]
    \centering
    \includegraphics[width=1\linewidth]{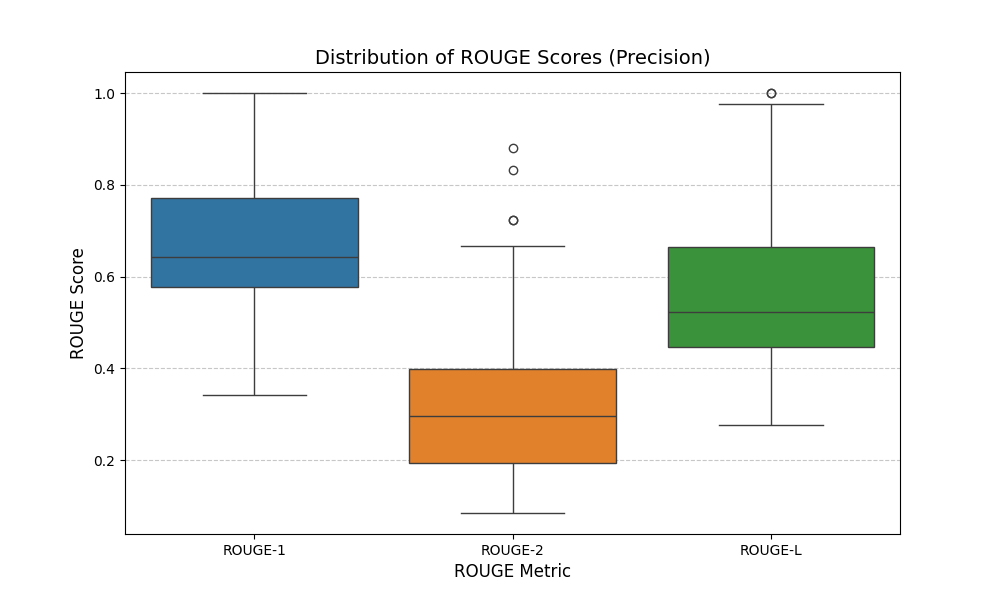}
    \caption{Distribution of ROUGE Scores for Precision}
    \label{precision}
\end{figure}

\textbf{Precision} measures how accurate the AI-generated summary is by checking how much of its content is actually relevant compared to the reference summary. A high precision means the AI isn’t adding unnecessary or unrelated details. Results of the precision analysis are in figure \ref{precision}.

\begin{figure}[h]
    \centering
    \includegraphics[width=1\linewidth]{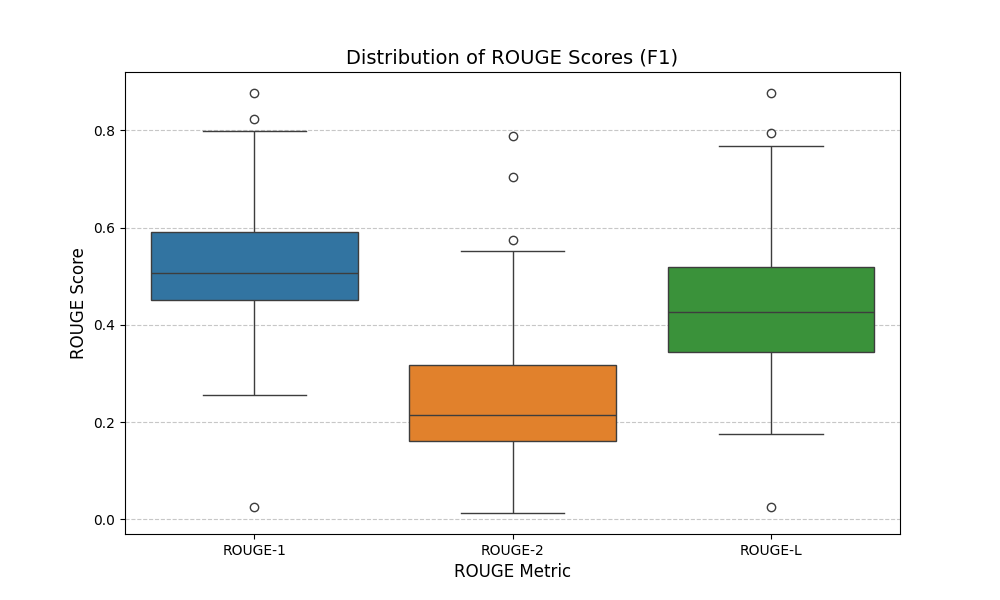}
    \caption{Distribution of ROUGE Scores for F1-Scores}
    \label{f1}
\end{figure}

\textbf{F1-score} is the balance between precision and recall. If precision is too high but recall is low, the AI is being too selective and missing important details. If recall is high but precision is low, the AI is including too much irrelevant information. F1-score finds the suitable spot between the two. The results of F1-score are illustrated in figure \ref{f1}.

\subsubsection{BERTscore Analysis}

BERTScore is computed using pre-trained Transformer models, specifically BERT (Bidirectional Encoder Representations from Transformers). It calculates similarity by:

\begin{enumerate}
    \item Tokenizing both reference and AI-generated summaries into word embeddings using Google's BERT, a transformers model. \cite{bert-original}

    \item Comparing each word in the generated summary with the most similar word in the reference summary based on cosine similarity.
    
    \item Aggregating the similarity scores across all tokens to produce an overall BERTScore, which includes precision, recall, and F1-score.

\end{enumerate}

\begin{figure}[h!]
    \centering
    \includegraphics[width=1\linewidth]{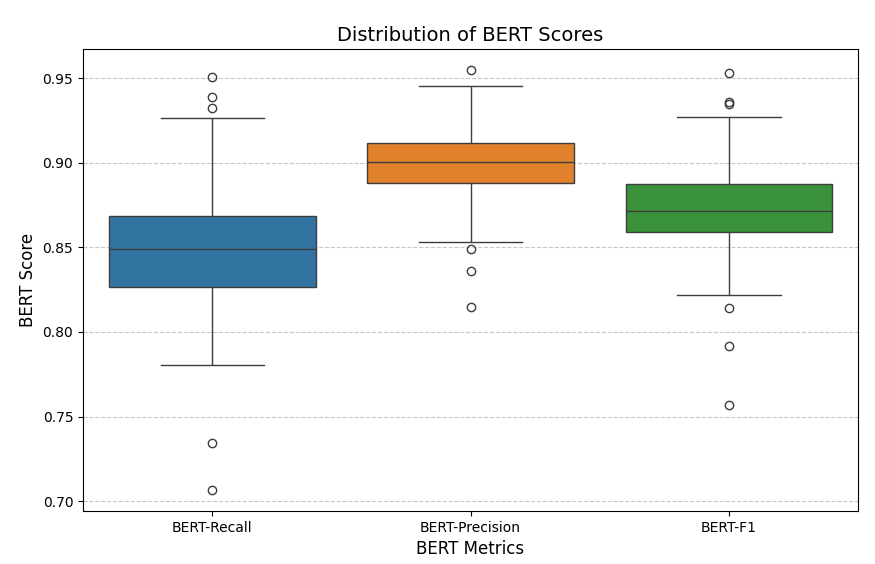}
    \caption{Distribution of BERTScore}
    \label{bert-score}
\end{figure}

BERTScore evaluates summaries using deep-learning-based embeddings. This allows for a more flexible comparison, considering synonyms and contextual meanings. BERTScore is valuable in medical summarization because it captures semantic similarity, ensuring that meaning is preserved even if wording differs. In addition, it helps assess how well the AI understands medical terminology and paraphrases effectively.\\
The same generated dataset that was used in calculating the ROUGE scores is used here: 150 reference medical histories and 150 generated medical summaries. The result of the BERTScore analysis are illustrated in figure \ref{bert-score}.

\subsection{Chatbot Results}

\subsubsection{Example of Chat Interaction}
The following figure, \ref{int}, shows the results of Llama model answer for a doctor, who initiates a session with: "What are the symptoms of Asthma?".

\begin{figure}[h!]
    \centering
    \includegraphics[width=1\linewidth]{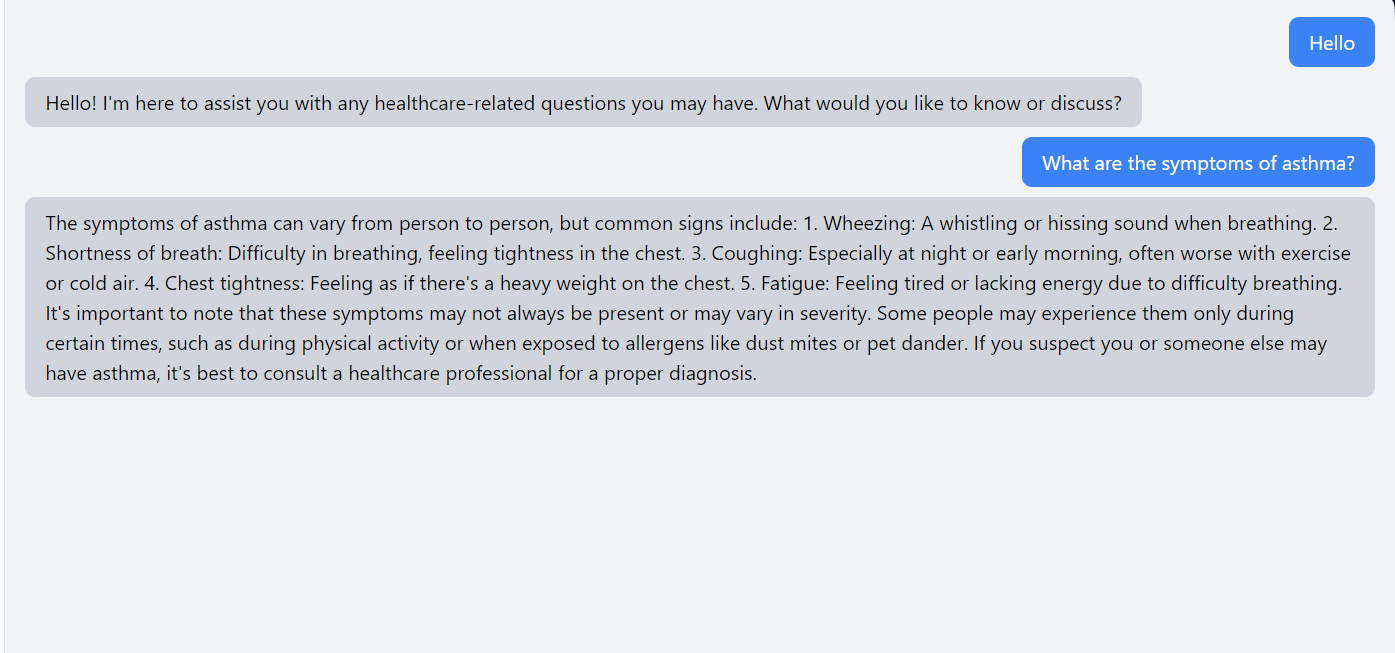}
    \caption{Chatbot initiation result}
    \label{int}
\end{figure}
\begin{figure}[h]
    \centering
    \includegraphics[width=01\linewidth]{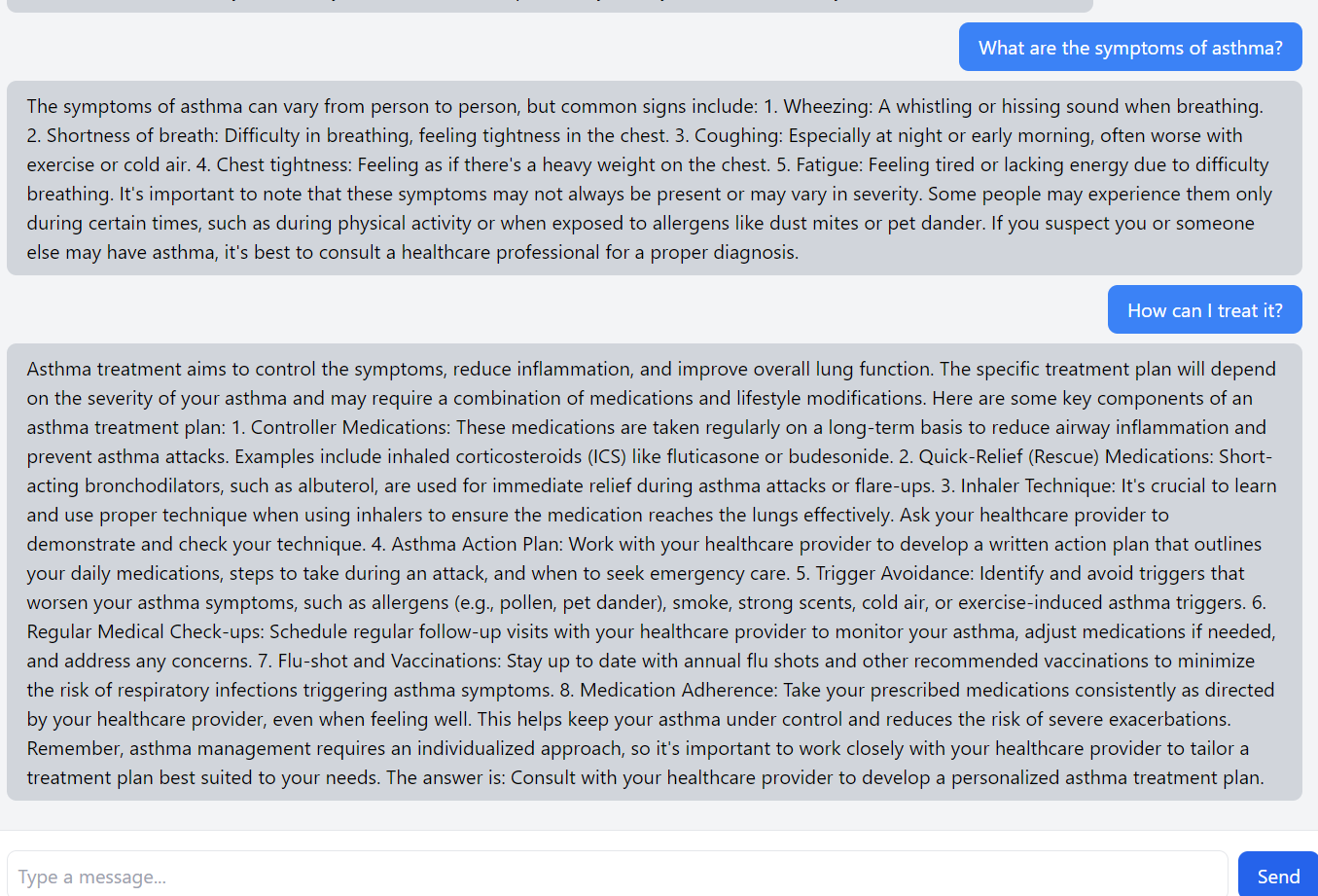}
    \caption{continues the conversation for chatbot}
    \label{fig:enter-label}
\end{figure}

\vspace{5cm}
 The doctor continues the conversation to ask follow-up questions: “How can I treat it?”, shown in figure \ref{fig:enter-label}
 Llama provides the answer then it is send back to the front-end using the back-end, while being saved back in MongoDB.

\subsubsection{Patient History Summarization Results}
\begin{figure}[h]
    \centering
    \includegraphics[width=1\linewidth]{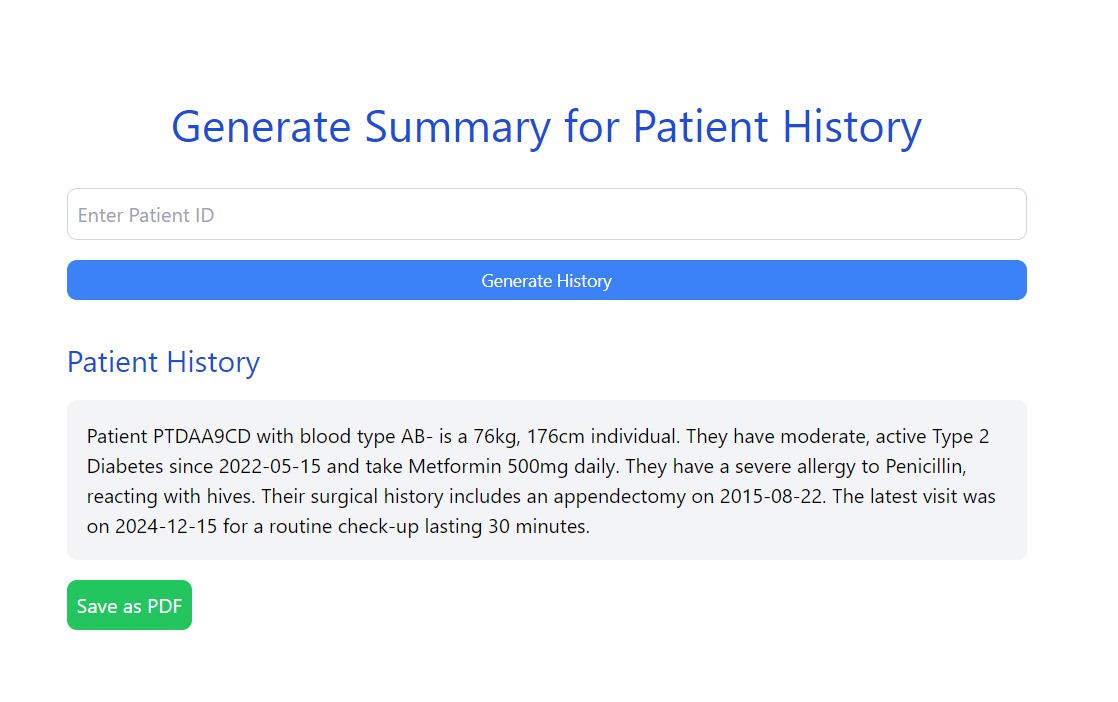}
    \caption{AI summarization tool result}
    \label{summ}
\end{figure}

Figure \ref{summ} represents the interaction between the doctor and summarization model. After sending patient ID, the model responds with a brief summary about the patient's history, highlighting the important data that a doctor would need before examination.
\vspace{5cm}

\subsubsection{Xray model results}
\begin{figure}[h]
    \centering
    \includegraphics[width=1\linewidth]{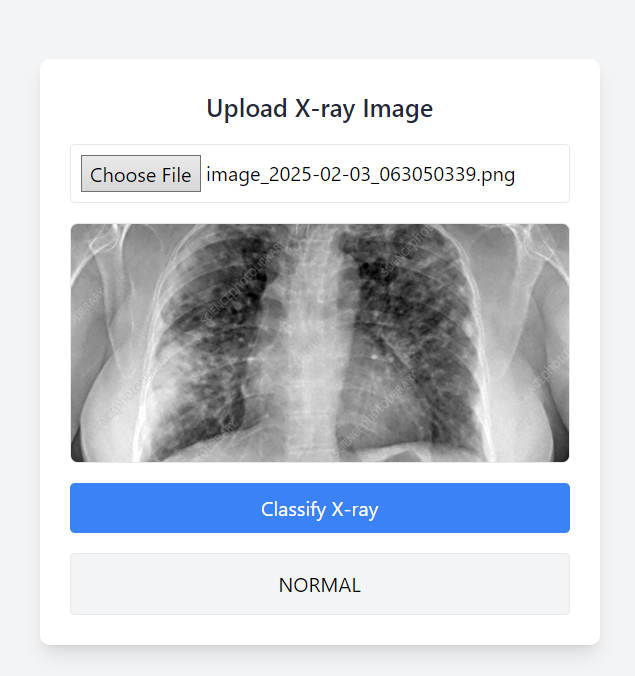}
    \caption{X-ray image of a healthy lung}
    \label{healthy}
\end{figure}

\begin{figure}[h]
    \centering
    \includegraphics[width=1\linewidth]{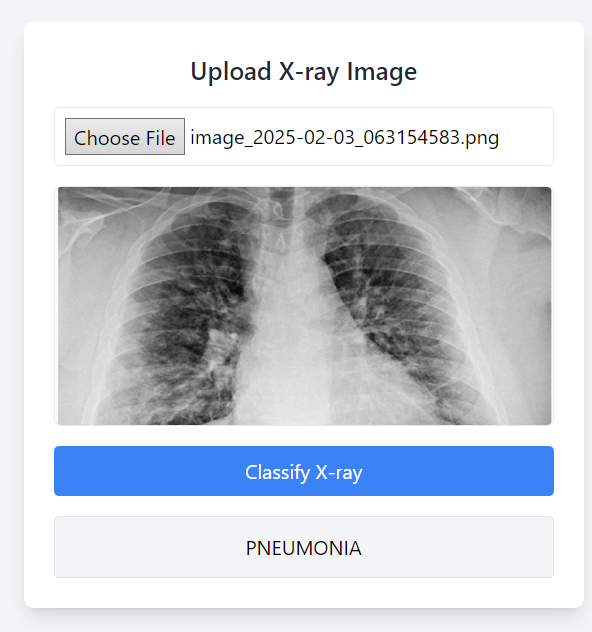}
    \caption{X-ray image of a lung with pneumonia.}
    \label{bad}
\end{figure}

This section includes AI model's analysis for two X-ray images of the lungs, each representing distinct conditions. The first image \ref{healthy} is a healthy lung. 
In contrast, the second image \ref{bad} illustrates a lung affected by pneumonia. 
The AI model's ability to differentiate between these conditions is demonstrated through its analysis, highlighting its potential as a diagnostic tool in medical imaging. These results underscore the model's capability to assist healthcare providers in identifying pathological changes, thereby enhancing diagnostic accuracy and efficiency.

\subsubsection{Performance Testing Results}
\begin{figure}[h]
    \centering
    \includegraphics[width=\linewidth]{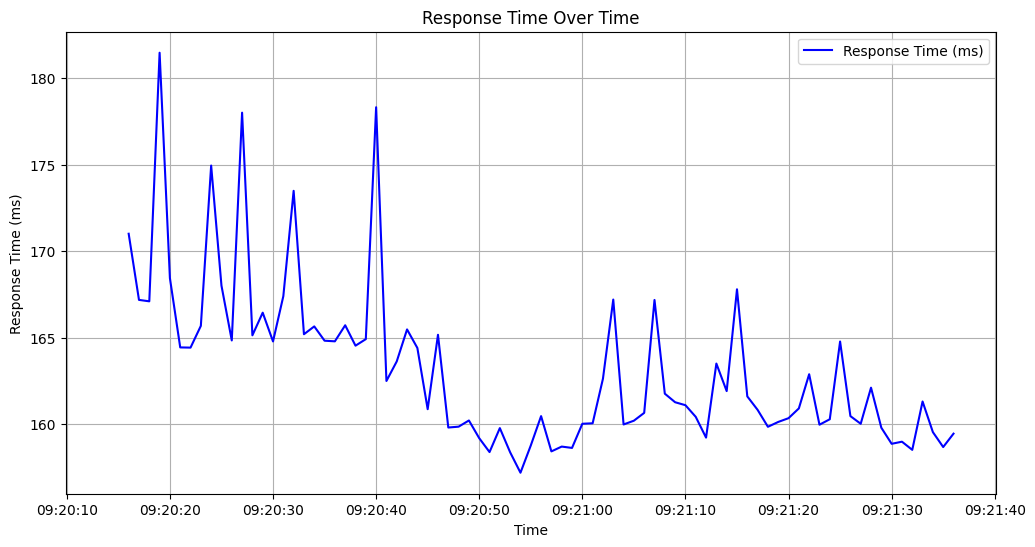}
    \caption{Response delay over time}
    \label{fig:res_over_time}
\end{figure}

The graphs illustrate the following key findings:

\textbf{\textit{Response Time Fluctuation:}} As seen in figure \ref{fig:res_over_time}, the response time exhibited fluctuations throughout the test, indicating variability in server performance. The average response time was observed to be within an acceptable range. However, there were a number of peaks showing response time increasing significantly above average, suggesting temporary performance bottlenecks during periods of higher load. Further investigation is necessary to determine the root causes of these fluctuations.

\begin{figure}[h]
    \centering
    \includegraphics[width=\linewidth]{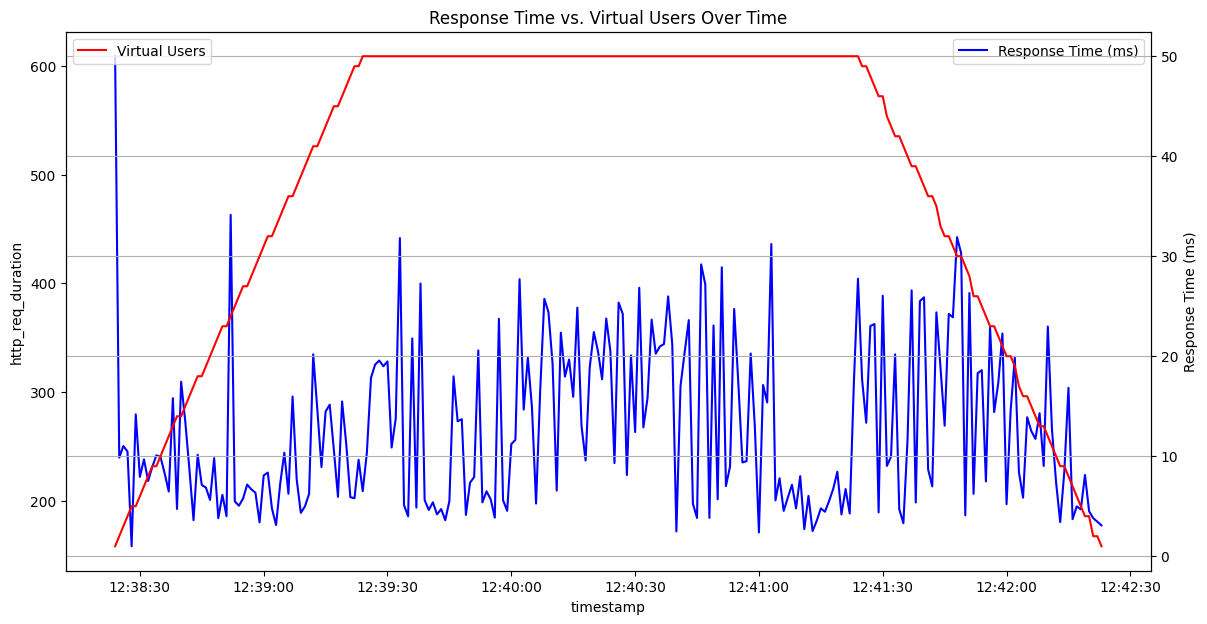}
    \caption{Response Time vs. Virtual Users over Time}
    \label{fig:vu_res_delay}
\end{figure}

\textbf{\textit{Correlation between Virtual Users and Response Time:}} Analysis of the (figure \ref{fig:vu_res_delay}) reveals a correlation between the number of virtual users and response time. As the number of VUs increased, there was a corresponding increase in response times. This indicates the application's performance might not scale linearly with increasing load.

\begin{figure}[h]
    \centering
    \includegraphics[width=\linewidth]{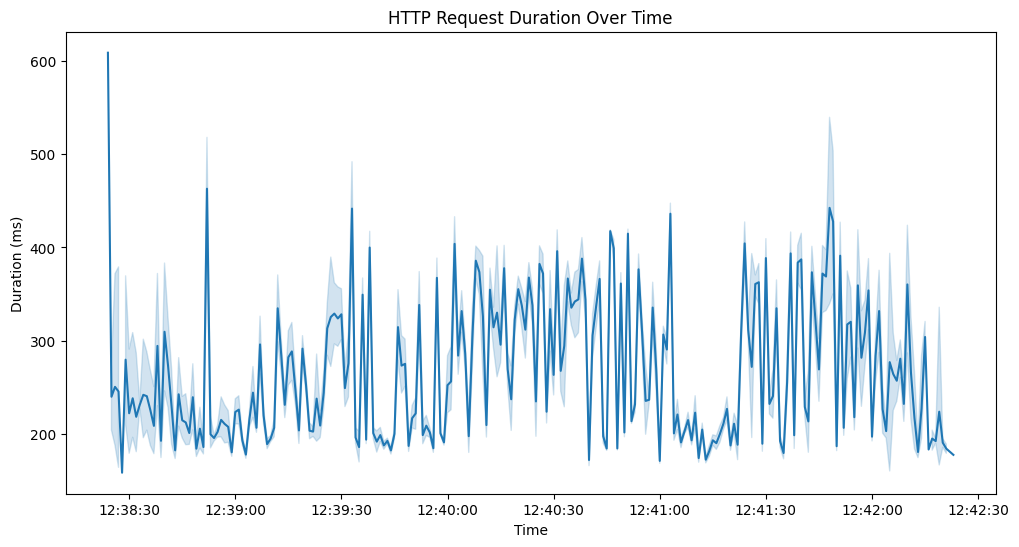}
    \caption{HTTP Request Duration Over Time}
    \label{fig:http_req_dur}
\end{figure}

\textbf{\textit{HTTP Request Duration:}} Figure \ref{fig:http_req_dur} shows variations in the duration of HTTP requests, including the time taken for request blocking, connecting, TLS handshaking, sending the request, waiting for a response, and receiving the response. The graph displays some peaks in request duration throughout the test, which could indicate intermittent performance bottlenecks or other underlying performance issues.

\section{Discussion}
\subsection{Survey Results Interpretation}
\begin{figure}[h]
                \centering
                \includegraphics[width=1\linewidth]{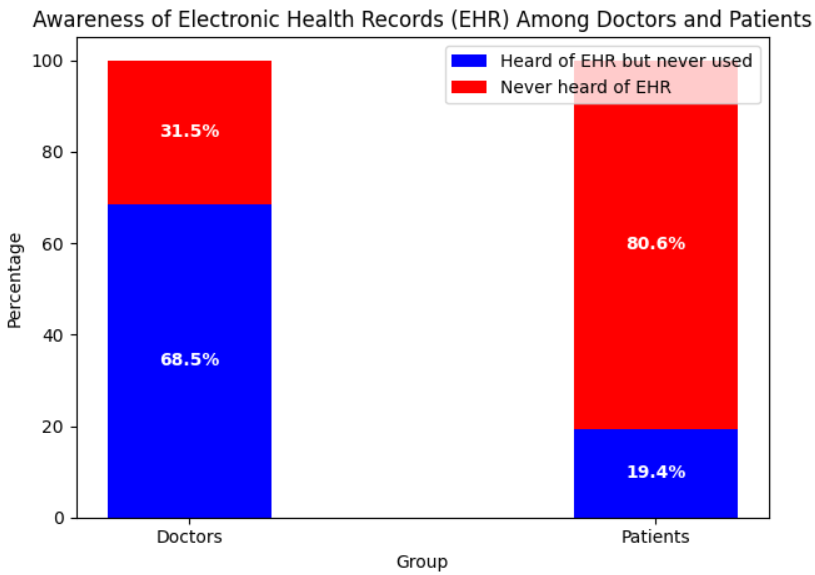}
                \caption{Graph of EHR Awareness.}
                \label{EHR Awareness}
\end{figure}

According to the survey results, there was a general lack of awareness about electronic health records (EHR), especially among patients as shown in figure~\ref{EHR Awareness} . Most participants acknowledged that the current paper-based system needed to be changed into an electronic one. There was hesitation about adopting technology due to security, technology barriers and data protection concerns. To address these worries, the EHR project is designed with a user- friendly interface to simplify the transition from a paper-based system to a digital one. 
            
Additionally, a multi-layered security approach is adopted to protect patient data, prevent unauthorized access, and ensure compliance with relevant standards such as HIPAA and GDPR. This approach includes encryption, multi-factor authentication, and regular security audits to maintain the highest level of data integrity and confidentiality. Furthermore, a chatbot was developed to assist doctors in their daily tasks, saving them time and improving efficiency. The chatbot also helps patients by providing instant responses to common queries, scheduling appointments, and offering reminders for medication or follow-up visits, thereby enhancing the overall patient experience and engagement. More than half of the participants expressed a lack of trust in using artificial intelligence tools for making medical decisions. Therefore, AI-tools were implemented to aid doctors for fasters examination; however, it is ensured that the final decision-making authority is given to the doctor. The system provides summarizing history of the patient and reading for X-rays, but the doctor must approve or reject them, ensuring human oversight and control.\\
It was observed that the functionality that received the most attention was keeping and showing medical historical features. Participants ensured the significance of having easy and secure access to the medical. Thus, this feature was prioritized in the new system, securing that all relevant medical information is available for both doctors and patients for the fast data retrieval.

\subsection{AI Results Interpretation}

\subsubsection{ROUGE Score Interpretation}

ROUGE score evaluates the lexical overlap between the AI-generated summaries and the reference medical summaries. As per the results in figures \ref{recall}, \ref{precision}, and \ref{f1} for the ROUGE scores, the analysis of it is as follows:

\textbf{ROUGE-1 (Unigram Overlap):}

\begin{itemize}
    \item The median ROUGE-1 score is relatively high. This indicates that the AI-generated summaries contain a strong overlap with the reference summaries at the word level.
    \item The interquartile range (IQR) is moderate. However, it shows some variations in performance across different cases.
    \item Outliers with lower scores suggest some summaries may miss important keywords.
\end{itemize}

\textbf{ROUGE-2 (Bigram Overlap):}
\begin{itemize}
    \item The lower median score here compared to ROUGE-1 suggests that while individual words match, the AI struggles to reproduce consecutive word pairs. This is critical for maintaining medical context.
    \item The wider spread of values indicates inconsistency in phrase formation.
\end{itemize}

\textbf{ROUGE-L (Longest Common Subsequence Overlap):}

\begin{itemize}
    \item The performance of ROUGE-L is between ROUGE-1 and ROUGE-2, indicating that the generated summaries maintain a reasonable structure but are not fully aligned with human-written references in terms of fluency.
    \item The distribution shows that while most summaries follow a structured pattern, some lose coherence in longer sequences.

\end{itemize}

\textbf{Precision, Recall, and F1-Score Analysis:}
\begin{itemize}
    \item Recall is generally higher than precision, meaning that the model captures more of the original content but may introduce redundancy.
    \item Precision is lower in ROUGE-2, showing that the AI model sometimes generates extra words that don’t precisely match the reference.
    \item F1-scores provide a balanced measure but highlight areas where improvement is needed in maintaining both completeness and accuracy.
\end{itemize}

\subsubsection{BERTScore Interpretation}

Unlike ROUGE, BERTScore uses deep-learning embeddings to measure semantic similarity rather than exact word matching. The BERTScore results indicate:

\textbf{Higher Scores Compared to ROUGE:}

\begin{itemize}
    \item The median BERTScore (F1) is above 0.85, demonstrating that even when exact words don’t match, the AI-generated summaries convey the same meaning.
    \item This is particularly useful in medical summarization, where paraphrasing is common.
\end{itemize}
\textbf{Precision vs. Recall in BERTScore:}

\begin{itemize}
    \item BERT-Precision is higher than ROUGE-Precision, meaning the model effectively selects relevant words from the medical history when generating summaries.
    \item BERT-Recall is slightly lower than its precision, suggesting that while the model includes relevant details, some aspects of the reference summaries might be missing.
\end{itemize}

\textbf{Consistency and Outliers:}
\begin{itemize}
    \item The narrower interquartile range (IQR) in BERTScore compared to ROUGE suggests more consistent performance across different summaries.
    \item Outliers represent cases where the AI-generated summary deviates significantly from the reference, potentially due to missing key medical details or rewording complex phrases.
\end{itemize}

\subsubsection{System Performance}
\begin{figure}[h]
    \centering
    \includegraphics[width=\linewidth]{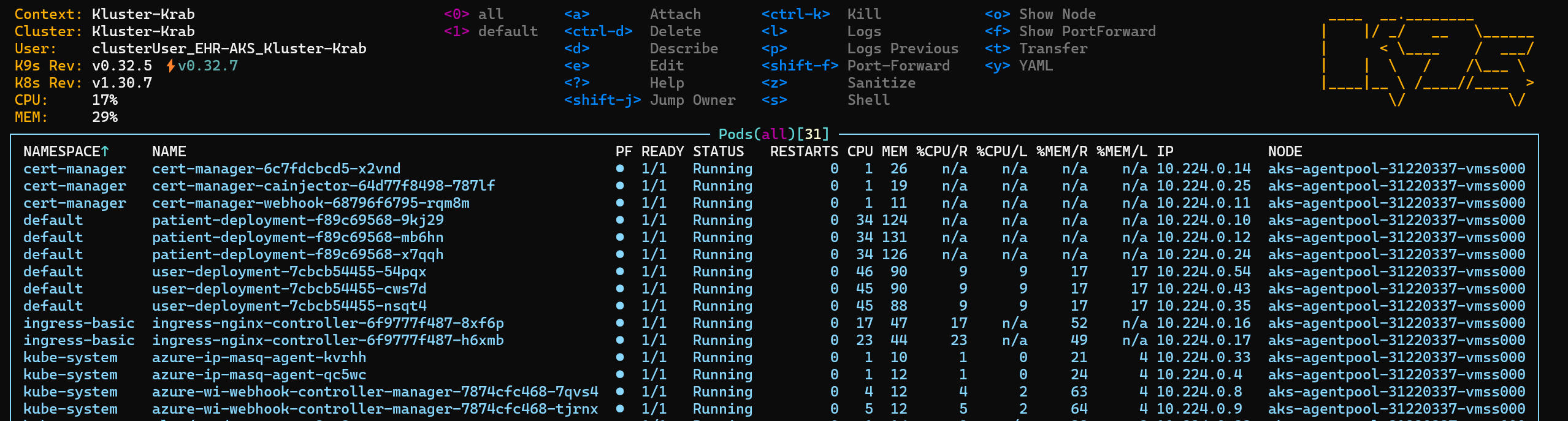}
    \caption{Number of Pods increase when under load}
    \label{fig:k9s}
\end{figure}

The performance testing results validate the effectiveness of the chosen architecture and implementation. The application demonstrated excellent performance, scalability, and stability under realistic user loads. These positive results can be directly attributed to the Kubernetes-based architecture, which allowed the application to dynamically scale resources based on demand as seen in Figure \ref{fig:k9s} where both microservices scaled to three instances under load.

The consistently low response times, even during periods of increased load, are particularly noteworthy. This indicates the Kubernetes auto-scaling mechanism functioned as expected, seamlessly provisioning additional resources to accommodate the growing number of virtual users. The steady increase in users did not adversely affect the application's performance, further demonstrating the resilience of the system. The results provide strong confidence in the application's ability to handle real-world user traffic and deliver a consistently responsive and reliable user experience. This robust performance foundation positions the application well for future growth and increased user demand.

\section{Conclusion}

To sum up, the survey responses indicate strong agreement in the need for transitioning from paper-based records to an EHR system, despite concerns about technological barriers, data security, and AI reliability. While medical professionals realize how much AI will help them, many of them are skeptical about fully integrating AI into clinical decision-making. On the other hand, Patients are interested but cautious which shows that privacy concerns and usability challenges are important. These provided valuable guidance while designing an AI-integrated, secure, and user-friendly EHR system to meet the expectations of both healthcare professionals and their patients.

This project successfully developed an AI-driven Electronic Health Record (EHR) system tailored for the Egyptian healthcare landscape, with a scalable backend architecture. The design of the system was centered around a microservices-based architecture that split the application into smaller, services responsible for exclusive functionalities such as user management, patient records, and artificial intelligence integration. This architecture allowed individual services to grow and evolve on their own and thus made the system more stable and easier to maintain. Data is stored and handled using a polyglot persistence strategy, with PostgreSQL used for relational data management of user administration and transactions, and MongoDB used to store flexible, document-based patient records and artificial intelligence results. This use of multiple database technologies improves both storing and retrieving different kinds of information and allows the backend to process complex medical information effectively. The use of a caching layer with Redis greatly enhances performance by reducing the workload on the database for data frequently accessed, such as ongoing user sessions, laboratory results, and artificial intelligence-generated summaries.

In addition, the backend boasts a comprehensive security system to protect sensitive patient information while maintaining compliance with relevant regulations, such as the Egyptian Data Protection Law. It consists of a robust end-to-end security strategy for every step of the request lifecycle, including authentication through Auth0, role-based access control, and strict data validation workflows. Machine-to-machine authentication ensures that only trusted microservices can talk to each other. Implementation is fully automated using GitHub Actions, enabling containerization through Docker and orchestration using Azure Kubernetes Service (AKS). This enables consistent and reliable deployment across environments, as well as scalable resource management through tailored node pools and dynamic adjustment based on demand.

The use of this architectural design, supported by efficient technologies like Express.js, Node.js, and Python with FastAPI for artificial intelligence modules, has enabled the creation of a robust and scalable backend to handle large volumes of patient data and user requests. The careful technology stack design and deployment process ensure the system's reliability, performance, and compliance with regulations and security, thus paving the way for a paradigm shift in overhauling Egypt's healthcare system. In addition, the use of structured but adaptable databases together with a microservices architecture supports possible future development of the system as well as integration with other healthcare information technology.

According to the ROUGE and BERTScore results, the AI integration has met its desired goal of the EHR system. BERTScore consistently outperformed ROUGE in assessing medical summaries since it captures semantic meaning rather than relying only on word-level overlap. Also, the higher recall observed across ROUGE and BERTScore indicates the prioritization of completeness. This is very advantageous in healthcare applications because all information is important and could impact the decisions made. However, this focus on recall can sometimes lead to longer summaries and may require further refinement for conciseness. Phrase continuity becomes problematic due to the poor ROUGE-2 scores. It shows that individual phrases are kept in the summary but the medical narrative structure has to be improved. Despite these difficulties, BERTScore performance is great with high precision and recall. Consequently, the model produces insightful summaries without adding undue noise. These observations highlight how crucial it is to apply semantic evaluation methods, such as BERTScore, to medical summarization jobs. Future optimization of the structure of clinical summaries produced by AI will benefit from this.

Additionally, the generation tasks, made by Llama3 OpenBioLLM, have provided automation to the doctor's repetitive tasks. Thus, doctors will have more time to care for the patient.

\section{Future Work}

\subsection{Integration with Other Healthcare Systems}
As the EHR system grows and becomes more integrated with Egypt's healthcare ecosystem, it'd be crucial to have the ability to share data seamlessly with other healthcare information systems. Effective communication ensures consistency, reduces redundancy and enhances collaboration between healthcare providers and medical institutions. In order to achieve this, the system must adopt the standardized protocols for health data exchange such as HL7 (Health level 7) and FHIR (Fast Healthcare Interoperability resources). These standards define a universal format for transmitting health-related information.

The current system uses REST APIs for internal microservices communication which provides interoperability within the system. However, when exchanging data with other existing systems specially hospitals, the system must adopt the industry standards for data communication. HL7v2 standards is the most widely used standards, it defines a message-based protocol for real-time data exchange, while FHIR, a more modern standard, leverages REST APIs and JSON/XML formats making it more suitable for cloud-based systems.

\subsubsection{Integration Process}
\begin{enumerate}
    \item An integration engine (Mirth Connect), will be deployed to act as a bridge between the EHR system and external systems. These engines are designed to parse, transform, and route HL7 messages, ensuring compatibility between different data formats.
    \item To ensure security, these connections are often established through a Virtual Private Network (VPN), as MLLP lacks native encryption. Alternatively, HL7v2 messages can be exchanged over web services or SFTP (Secure File Transfer Protocol), though these methods are less common.

    \item The integration engine will be configured to parse incoming HL7 messages, convert them into a format the EHR system can understand (e.g., JSON), and route them to the appropriate microservice. Similarly, outgoing data from the EHR system will be transformed into HL7 or FHIR-compliant messages before being sent to external systems.
\end{enumerate}

\begin{figure}[h]
    \centering
    \includegraphics[width=1\linewidth]{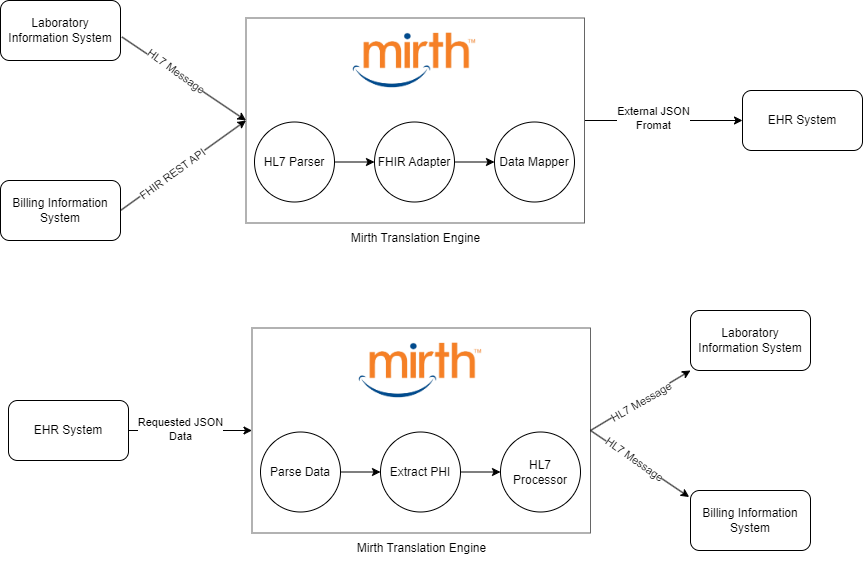}
    \caption{Mirth Connect Transaltion Architecture for Incoming and Outgoing Requests}
    \label{fig:Mirth Connect Transaltion Architecture}
\end{figure}

Figure \ref{fig:Mirth Connect Transaltion Architecture} shows how Mirth Connect enables communication with other healthcare information systems. A piece of information requested from the EHR system would first be passed to Mirth Connect which parses the information sent, extracts the relevant information, then the HL7/FHIR Processor would transform the extracted information into the appropriate format to be sent to the requesting entity. The same process can be used for sending information from the EHR system to others. 

\subsubsection{Advantages of Microservices for Interoperability}
The systems microservices architecture makes the process of integration with other healthcare information systems simple. It allows the decoupling of the core functionalities of the system from the translation logic. A dedicated translation engine would be developed to handle the process of transforming data from it's original format to another compatible with HL7 and FHIR. this modular approach allows the system to adapt to new standards without affecting the existing system. An example of a translation engine is Mirth Connect. It's an open-source integration engine specialized in healthcare.

\subsection{AI Summarization System Improvement}
While the AI-driven medical summarization system demonstrates promising results, several areas remain for further exploration and enhancement. Future work should focus on refining the summarization model, introduction of Retrieval-Augmented Generation (RAG) system, and addition of multiple disease classification models. Key directions for future research include:

\subsubsection{Introduction of Retrieval-Augmented Generation}

One of the main limitations of the current approach is that the AI model generates summaries based on input medical history without accessing external medical knowledge. Retrieval-Augmented Generation (RAG) can enhance summarization by incorporating relevant external medical databases (e.g., PubMed, SNOMED CT, or clinical guidelines) to enrich responses with up-to-date. In addition, the RAG system will help the AI model to navigate through the database to get the patient data, scans, and lab data. This ensures that summaries are both accurate and clinically relevant.

\subsubsection{Fine-tuning the OpenBioLLM on Egyptian Medical Data}

Even though Llama3 OpenBioLLM is fine-tuned on the state-of-the-art datasets form multiple global and certified sources, there should be focus on Egypt and its medical necessities. The proposed EHR will be centralized across the Egyptian Public Hospitals and the model should be well-aware about the current and the previous health trends in Egypt. Some diseases are more common here in Egypt than other parts of the world. Also, training the model on Egyptian medical history datasets will be very efficient in increasing the accuracy of the model. Lastly, Arabic language should be supported by the model for conversation via the chatbot.

\subsubsection{Expanding the System to Support Multiple Disease Classification Models}

To enhance diagnostic capabilities, the system should extend beyond summarization and incorporate automated disease classification using disease detection models. Since the Vision Transformer (ViT) Pneomnia Classificaion model has proved its accuarcy, it could be used for more radiological diagnostics. Lastly, more models could be added to the system to automate the disease detection.

\section*{Acknowledgment}

The authors would like to express their gratitude to Dr. Ahmed Fares their supervisor. Further, appreciation is extended to Dell Technologies' AI Empower Egypt program for their sponsorship of this project.

\bibliographystyle{IEEEtran}
\bibliography{bibliography}

\newpage
\appendices
\onecolumn
\setlength{\parskip}{1pt plus 1pt minus 1pt}

\begin{singlespace}
\section{Further Discussion on \textbf{patient-service} ERD}
This diagram represents the complete ERD for the \textbf{patient-service}, illustrating all entities and their relationships within the system.
\end{singlespace}

\begin{figure*}[h]
    \centering
    \includegraphics[width=\textwidth]{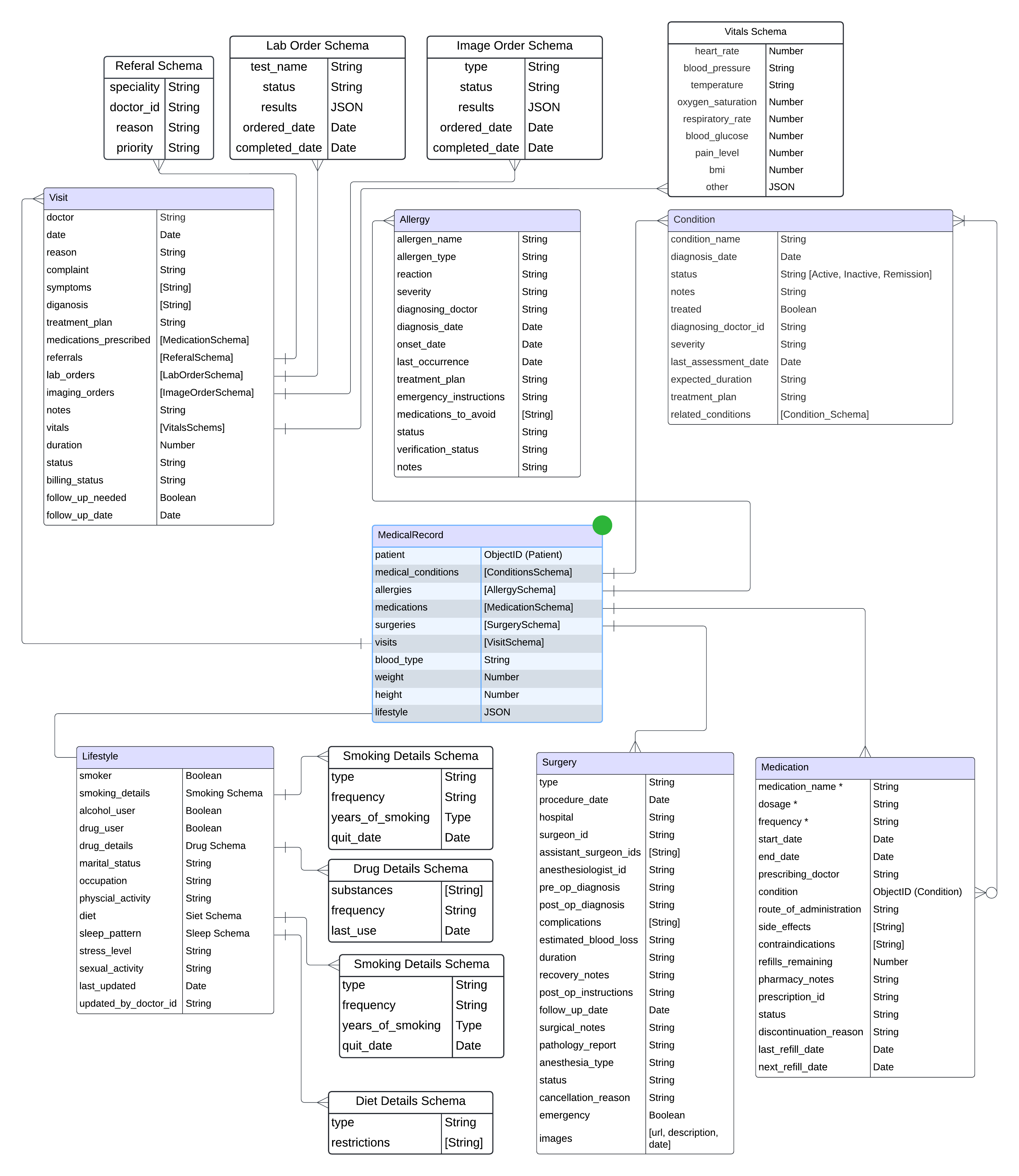}
    \caption{ERD for the patient-service}
    \label{fig:patient-service}
\end{figure*}

\newpage

\setlength{\parskip}{1pt plus 1pt minus 1pt}

\begin{singlespace}
\section{Further Discussion on \textbf{EHR System Interface}}
This use-case diagram represents the complete cycle for the actors (admin, patient, doctor), showing their interactions and functionalities.
\end{singlespace}

\begin{figure*}[h]
    \centering
    \includegraphics[width=0.98\textwidth]{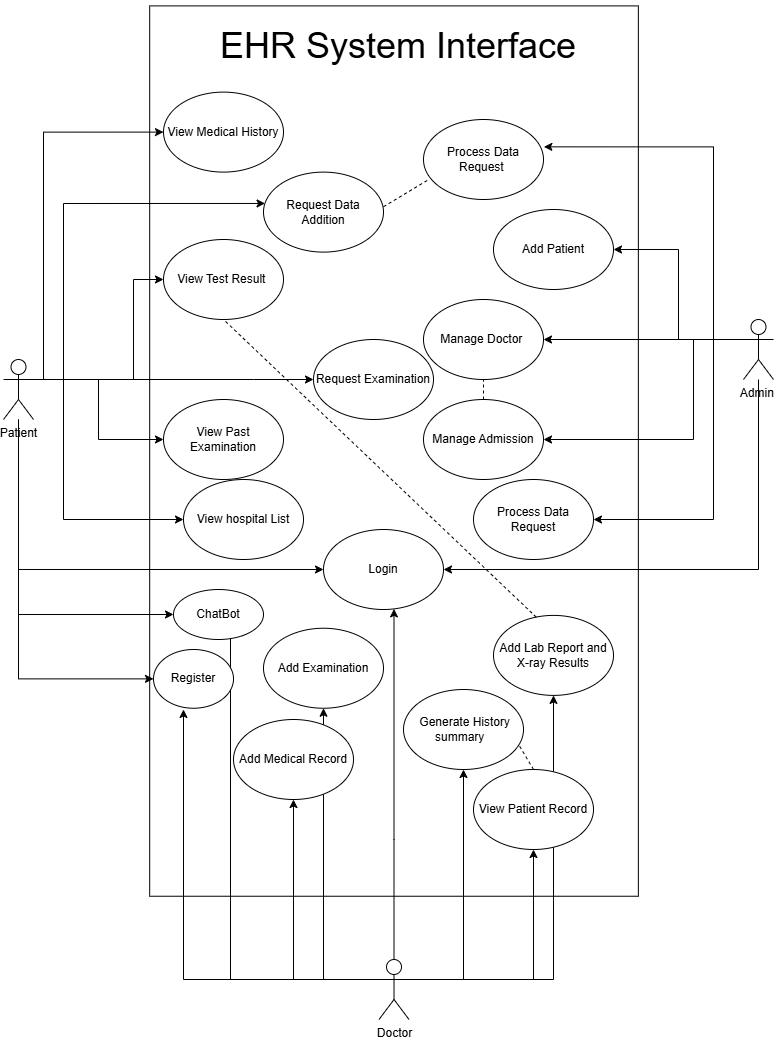}
    \caption{Use Case Diagram of User Interface}
    \label{fig:use-case}
\end{figure*}
\end{document}